\documentclass{aa}
\usepackage{longtable}
\usepackage{graphicx}
\usepackage{txfonts}
\begin{document}
  \title{Molecular jets driven by high-mass protostars: a detailed study of the IRAS\,20126+4104 jet
  \thanks{Based on observations collected at UKIRT, TNG, and at Subaru and ISO data archive.}
       }

   \author{A. Caratti o Garatti \inst{1},
   	   D. Froebrich \inst{2},
	   J. Eisl\"{o}ffel \inst{1},	   
           T. Giannini \inst{3},
	  \and
	   B. Nisini \inst{3}
          }

   \offprints{A. Caratti o Garatti \email{caratti@tls-tautenburg.de}}

   \institute{Th\"uringer Landessternwarte Tautenburg,
              Sternwarte 5, D-07778 Tautenburg, Germany\\
              \email{caratti;jochen@tls-tautenburg.de}
              \and
	       Centre for Astrophysics and Planetary Science, University of Kent, Canterbury, CT2 7NH, United Kingdom\\
	      \email{df@star.kent.ac.uk}
	      \and
             INAF - Osservatorio Astronomico di Roma, via Frascati 33, I-00040 Monte Porzio, Italy\\
             \email{giannini;nisini@oa-roma.inaf.it}
             }

\date{Received; accepted }


  \abstract
   {Protostellar jets from intermediate- and high-mass protostars provide an excellent opportunity to understand the mechanisms responsible for intermediate- and high-mass star formation. A crucial question is if they are scaled-up versions of their low-mass counterparts.
Such high-mass jets are relatively rare and, usually, they are distant and highly embedded in their parental clouds. The 
IRAS\,20126+4104 molecular jet, driven by a 10$^4$\,$L_{\sun}$ protostar, represents a suitable target to investigate.}
   {We present here an extensive analysis of this protostellar jet, deriving the kinematical, dynamical, and physical conditions of the H$_2$ gas along the flow. }
  {The jet has been investigated by means of near-IR H$_2$ and [\ion{Fe}{ii}] narrow-band imaging, high resolution spectroscopy of the 1-0\,S(1) line (2.12\,$\mu$m), NIR (0.9-2.5\,$\mu$m) low resolution spectroscopy, along with ISO-SWS and LWS spectra (from 2.4 to 200\,$\mu$m). }
  {The flow shows a complex morphology. In addition to the large-scale jet precession presented in previous studies, 
we detect a small-scale wiggling close to the source, that may indicate the presence of a multiple system. The peak radial velocities of the H$_2$ knots range from -42 to -14\,km\,s$^{-1}$ in the blue lobe, and from -8 to 47\,km\,s$^{-1}$ in the red lobe.
  The low resolution spectra are rich in H$_2$ emission, and relatively faint [\ion{Fe}{ii}] (NIR), [\ion{O}{i}] and [\ion{C}{ii}] (FIR) emission is observed in the region close to the source. A warm H$_2$ gas component has an average excitation temperature that ranges between 2000\,K and 2500\,K. Additionally, the ISO-SWS spectrum reveals the presence of a cold component (520\,K), that strongly contributes to the radiative cooling of the flow and plays a major role in the dynamics of the flow.
  The estimated $L_{H_2}$ of the jet is 8.2$\pm$0.7\,$L_{\sun}$, suggesting that IRAS\,20126+4104 has an accretion rate 
significantly increased compared to low-mass YSOs. This is also supported by the derived mass flux rate from the H$_2$ lines ($\dot{M}_{out}$(H$_2$)$\sim$7.5$\times$10$^{-4}$\,$M_{\sun}$\,yr$^{-1}$).
The comparison between the H$_2$ and the outflow parameters strongly indicates that the jet is driving, at least partially, the outflow. As already found for low-mass protostellar jets, the measured H$_2$ outflow luminosity is tightly related to the source bolometric luminosity. }   
{As for few other intermediate- and high-mass protostellar jets in the literature, we conclude that 
IRAS\,20126+4104 jet is a scaled-up version of low-mass protostellar counterparts. }
\authorrunning{A.Caratti o Garatti et al.}
\titlerunning{The IRAS\,20126+4104 molecular jet}
\maketitle

\section{Introduction}
\label{intro:sec}

Protostellar jets and outflows are a ubiquitous phenomenon among young stellar objects (YSOs) of different mass and luminosity (see e.\,g. Shepherd~\cite{shepherd03}). They are usually explained as a consequence of accretion from a disc around the protostar (see e.\,g. Pudritz \& Norman~\cite{pudritz}, Camenzind~\cite{camenzind}).
This is particularly true for low- and, partially, for intermediate- and high-mass YSOs up to $L_{bol}\sim$10$^4$\,$L_{\sun}$ (or spectral type B0), where collimated outflows, often driven by protostellar jets, have been observed (Shepherd~\cite{shepherd03}). 
On the contrary, no highly collimated outflow or circumstellar disc have been observed in high-mass protostars exceeding 10$^5$\,$L_{\sun}$ 
(or O-type stars, the spectral type and $L_{bol}$ of the M17 disc silhouette are not clear yet) (see e.\,g. Arce et al.~\cite{arce}, Zinnecker \& Yorke~\cite{zinn}), and the formation mechanism of these latter objects is still debated. 
The observations are, however, strongly limited by the large distance of the massive star forming regions, the considerable extinction, and the short lifetime of massive YSOs. In addition, these objects are often grouped in small clusters, which confuses the morphology of massive star forming regions even more.
Therefore, optical and IR studies of intermediate- high-mass protostellar jets are very rare, and only a few examples are present in literature: e.\,g. IRAS\,18162-2048 ($L_{bol}\sim$2$\times$10$^4$\,$L_{\sun}$, Mart\'{i} et al.~\cite{marti}),
IRAS\,20126+4104 ($L_{bol}\sim$10$^4$\,$L_{\sun}$, Ayala et al.~\cite{ayala}), IRAS\,16547-4247 (6$\times$10$^4$\,$L_{\sun}$, Brooks et al.~\cite{brooks}), IRAS\,18151-1208 (2$\times$10$^4$\,$L_{\sun}$, Davis et al.~\cite{davis04}), and IRAS\,11101-5829 (10$^4$\,$L_{\sun}$, Gredel~\cite{gredel}), the M17 disc silhouette ($M_{YSO}\sim$15\,$M_{\sun}$, N\"{u}rnberger et al.~\cite{nur}).
Expanding the number of observations of intermediate- and high-mass jets and comparing their general properties 
with those of low-mass protostellar jets is therefore important to understand if differences exist, or if
high-mass protostellar jets are just scaled up versions of their low-mass counterparts.
We therefore investigate the kinematical and physical properties of the IRAS\,20126+4104 jet by means of NIR narrow-band
imaging, high resolution and low resolution IR spectroscopy. We then compare our findings with those of other high- and low-mass protostellar jets available in literature.

IRAS\,20126+4104, at a distance 1.7\,kpc, is a very-well studied high-mass YSO ($M\sim$7\,$M_{\sun}$, Cesaroni et al.~\cite{cesaroni97}, Cesaroni et al.~\cite{cesaroni05}), in a very early stage of evolution. It is accreting mass at a very high rate ($\dot{M}_{acc}\sim$2$\times$10$^{-3}$\,$M_{\sun}$yr$^{-1}$, Cesaroni et al.~\cite{cesaroni05}) and it gives birth to a large poorly-collimated CO outflow. It harbours the first H$_2$ jet detected from a high mass YSO (Ayala et al.~\cite{ayala}), which
had previously been seen in SiO emission close to the source (Cesaroni et al.~\cite{cesaroni99}).
The H$_2$ jet extends for about 1\,pc. Its `S' shape morphology suggests that it is precessing with a period of $\sim$60\,000\,yr and with a wide precession angle of about 37$\degr$ (Shepherd et al.~\cite{shepherd}). As a consequence, the inclination of the flow with respect to the plane of the sky changes strongly from $\sim$9$\degr$, close to the source, up to $\sim$45$\degr$ in the outer part of the flow.

The structure of this paper is as follows. In Sect.~\ref{observations:sec} our observations are presented. 
Sect.~\ref{results:sec} reports an overview of our results, including the
morphology of the H$_2$ jet, its kinematics, the physical parameters of the gas, its energy and mass flux rate. 
In Sect.~\ref{discussion:sec} we briefly consider the cause of the newly detected small-scale precession mode.
Then, we discuss our H$_2$ data in relation to the CO outflow literature data.
Finally, we compare the properties of the IRAS\,20126+4104 jet with other high- and low-mass protostellar jets.

\section{Observations and data reduction}
\label{observations:sec}

Our data were collected at the UK Infrared Telescope (UKIRT), and at the 3.5-m Italian Telescopio Nazionale Galileo (TNG).
Further data have been retrieved from the Subaru and ISO archives\footnote{These data are available at http://smoka.nao.ac.jp/ (Subaru),
and http://www.iso.vilspa.esa.es/ida/index.html (ISO)}. The relevant information on the observational settings is summarised in Table~\ref{obs:tab}.

\begin{figure}
\centering
\includegraphics [width=9.2 cm] {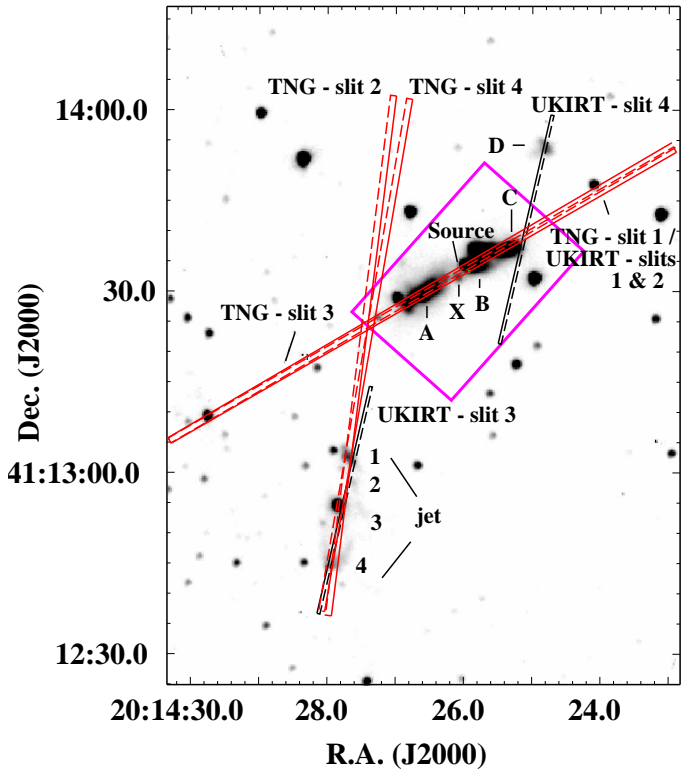}
   \caption{ Close up of the TNG H$_2$ mosaic, showing the entire flow with superimposed positions of the slits, and
the ISO-SWS FoV. The positions of the H$_2$ knots and the IRAS source ($\alpha=20^h14^m26.03^s$ and $\delta$ =41$\degr$13$\arcmin$32$\arcsec$.58 [J2000], Hofner et al.~\cite{hofner}) are also indicated.
\label{slits:fig}}
\end{figure}

\begin{table*}
\caption[]{ Journal of observations     \label{obs:tab}}
\begin{center}
IMAGING\\[+2pt]
\begin{tabular}{cccccc}
\hline \hline\\[-5pt]
Date of obs. &      Telescope/  &  Filter  & Resolution & seeing & Exp. Time      \\
(d,m,y)   &         Instrument  &  Band & ($\arcsec$/pixel) & ($\arcsec$) & (s) \\
\hline\\[-5pt]                                                                                                                       
01.10.2006  & UKIRT/UIST  & [\ion{Fe}{ii}] & 0.12 & 0.6 & 720  \\
01.10.2006  & UKIRT/UIST  &    H           & 0.12 & 0.6 & 360  \\
06.08.2006  & TNG/NICS    &  H$_2$         & 0.25 & 1.2 & 900 \\
06.08.2006  & TNG/NICS    &  K$^\prime$    & 0.25 & 1.2  & 90  \\
10.07.2003  & Subaru/CIAO    &  H$_2$, Br$\gamma$, K$_{cont}$    & 0.022 & 0.9  & 1800  \\
\hline\\[-5pt]
\end{tabular}\\
SPECTROSCOPY\\[+2pt]
\begin{tabular}{ccccccccc}
\hline \hline\\[-5pt]
 Date of obs & Telescope/  &.  Wavelength &  t$_{int}$ & P.A. & Slit/Aperture width & $\mathcal R$ & Encompassed Knots & Notes \\
 (d,m,y) &   Instrument  &  ($\mu$m) & (s) &  ($\degr$)  & ($\arcsec$) &  &  &\\
 \hline\\[-5pt]
01.10.2006 & UKIRT/CGS4    & 2.1218   & 2700 & 300   & 0.5 & 18\,500 & A1,A2,X,B,C2 & UKIRT-Slit1 \\
01.10.2006 & UKIRT/CGS4    & 2.1218   & 2700 & 300   & 0.5 & 18\,500 & A1,A2,B,C1 & UKIRT-Slit2 \\
01.10.2006 & UKIRT/CGS4    & 2.1218   & 2700 & 347   & 0.5 & 18\,500 & jet - knots 1,2,3,4 & UKIRT-Slit3 \\
01.10.2006 & UKIRT/CGS4    & 2.1218   & 2700 & 347   & 0.5 & 18\,500 & D,C2 & UKIRT-Slit4 \\
06.08.2006 & TNG/NICS      & 0.88--1.45   & 2000 & 299 & 1 & 500 & A,B,C & TNG-Slit1  \\
06.08.2006 & TNG/NICS      & 1.09--1.80   & 1800 & 299 & 1 & 500 & A,B,C & TNG-Slit1 \\
06.08.2006 & TNG/NICS      & 1.40--2.47   & 1800 & 299 & 1 & 500 & A,B,C & TNG-Slit1 \\
06.08.2006 & TNG/NICS      & 1.40--2.47   & 1800 & 350   & 1 & 500 & jet & TNG-Slit2  \\
18.10.2003 & TNG/NICS      & 1.09--1.80   & 2400 & 300 & 1 & 500 & A,B,C & TNG-Slit3 \\
18.10.2003 & TNG/NICS      & 1.40--2.47   & 2400 & 300 & 1 & 500 & A,B,C & TNG-Slit3 \\
18.10.2003 & TNG/NICS      & 1.09--1.80   & 1800 & 353.5 & 1 & 500 & jet & TNG-Slit4 \\
18.10.2003 & TNG/NICS      & 1.40--2.47   & 2400 & 353.5 & 1 & 500 & jet & TNG-Slit4 \\
11.05.1996 & ISO/SWS       & 2.38--45.2   & 1912 & $\cdots$ & 33$\times$20 & 1000--2000 & A,B,C,D &  \\
30.12.1995 & ISO/LWS       &  43--197   & 750 & $\cdots$ & 80 & 200 & A,B,C,D &  \\
\hline \hline
\end{tabular}
\end{center}
\end{table*}

\subsection{Imaging: H$_2$ \& [\ion{Fe}{ii}]}
\label{obs_ima:sec}

We used narrow-band filters centred on the H$_2$ (2.12\,$\mu$m) and [\ion{Fe}{ii}] (1.64\,$\mu$m) lines to detect both molecular
and ionic emission along the flow. Additional broad-band images ($H$ and $K^\prime$) were gathered to remove the continuum.
Our images were collected at UKIRT, using the near-IR instrument UIST (Ramsay Howat et al.~\cite{ramsay}) ([\ion{Fe}{ii}], H), and at the TNG, using NICS (Baffa et al.~\cite{baffa}) (H$_2$, $K^\prime$). All the raw data were reduced using IRAF\footnote{IRAF (Image Reduction and Analysis Facility) is distributed by the National Optical Astronomy Observatories, which are operated by AURA, Inc., cooperative agreement with the National Science Foundation.} packages, applying standard procedures for sky subtraction, dome
flat-fielding, bad pixel and cosmic ray removal, and image mosaicing. The resulting mosaics cover a region around IRAS\,20126+4104 of 3\farcm4$\times$3\farcm4 for UIST, and 5\farcm6$\times$5\farcm6 for NICS.
The calibration for both instruments was obtained by means of photometric standard stars observed in both narrow- and broad-band filters.
In the calibrated and continuum-subtracted narrow-band images, we measured H$_2$ and
[\ion{Fe}{ii}] fluxes for each detected knot using the task {\em polyphot} in IRAF, defining each region within a
$3 \sigma$ contour level above the sky background.

Additional high-resolution narrow-band images (H$_2$, Br$_{\gamma}$, and K$_{cont}$) of the central part of the molecular flow ($\sim$35$\arcsec\times$35$\arcsec$) were retrieved from the Subaru data archive (Baba et al.~\cite{baba}). The set of data were taken with the Coronagraphic Imager with Adaptive Optics (CIAO) (Murakawa et al.~\cite{murakawa}), used as a high-resolution NIR imager (0\farcs022/pixel). 
These data were used for a morphological study of the flow close to the source only (see Sect.~\ref{ima_res:sec} and \ref{prec:sec}).

Finally, we used K-band stars from the Two Micron All Sky Survey (2MASS) for the astrometric calibration of all the images.
As a result the formal errors for the plate solutions are $\sim$0\farcs5, 0\farcs4, and 0\farcs2, for the NICS, UIST, and CIAO images, respectively.

\subsection{Spectroscopy}

\subsubsection{Low resolution spectroscopy}

NIR low resolution spectroscopy ($\mathcal R\sim$500, slit width 1$\arcsec$) was acquired with NICS during two runs at the TNG, 
using three different grisms covering all the full NIR spectrum (0.9-2.5\,$\mu$m) (see Table~\ref{obs:tab}). The position of the slits is shown in Fig.~\ref{slits:fig}. Slits 1 and 3 are positioned on the knots close to the source (knots A, B, and C) and  parallel to the jet axis, covering almost all the H$_2$ emitting area. Slits 2 and 4 encompass the so called `jet' (see Ayala et al.~\cite{ayala}), that is the farthest H$_2$ emission in the red lobe (see also Fig.~\ref{slits:fig}). 

To perform our spectroscopic measurements, we adopted the usual ABB$'$A$'$ configuration, with a
total integration time between 1800\,s and 2400\,s per grism per slit. Each
observation was flat fielded, sky subtracted and corrected for the
curvature caused by longslit spectroscopy, while atmospheric
features were removed by dividing each spectrum by a telluric
standard star (spectral O type). The
wavelength calibration was obtained from Xenon and Argon lamps.
The flux was calibrated observing the photometric standard star AS\,35-0 (Hunt et al.~\cite{hunt}).
As the used grisms overlap between $\sim$1.1 and 1.8\,$\mu$m (see Tab.~\ref{obs:tab}), the single spectra
were combined to obtain deeper spectra. 
Finally, since no substantial difference was found between the pairs of spectra from Slit 1 and 3, nor from Slit 2 and 4, they 
were combined together to get a deeper spectrum of each observed knot.

\subsubsection{High resolution spectroscopy}

Our H$_2$ 1-0\,S(1) ($\lambda_{\rm vac} = 2.1218356~\mu$m; Bragg et al.~\cite{bragg}) echelle spectra were obtained at UKIRT using the spectrometer CGS4  (Mountain et al.~\cite{mountain}) (see Table~\ref{obs:tab}), equipped with a 256$\times $256 pixel InSb array (0\farcs41$\times$0\farcs90 pixel scale, in the dispersion and spatial direction, respectively). A 1-pixel-wide slit was used, resulting in a velocity resolution of 
$\sim$7.4\,km\,s$^{-1}$\,pixel$^{-1}$, and corresponding to a nominal resolving power of $\sim$18\,500. The instrumental profile in the dispersion direction, measured from Gaussian fits to the OH sky lines, was $\sim$10.9\,km\,s$^{-1}$ (or 1.46\,\AA). 

The targets and position angles (P.A.s) of the four slits are reported in Table~\ref{obs:tab}, while the slit positions are shown in Fig.~\ref{slits:fig} and encompass all the H$_2$ knots of the flow. Slit 1 and 2 are parallel to the flow axis and positioned on the knots close to the source (knots A, B, and C). Slit 3 encompasses the so-called `jet', and Slit 4 encompasses knot D and partially knot C.
The total exposure time for each position is 2700\,s.

The raw spectra were wavelength calibrated using the bright OH lines (Rousselot et al.~\cite{rousselot}) detected on each frame.
The employed IRAF tasks ({\em identify, re-identify, fitcoords}, and {\em transform}) also correct for the spatial distortion, and a
3rd-order fit in two dimensions has been used. As a result, the calibration is accurate to $\sim$3\,km\,s$^{-1}$.
The spectra were not flux calibrated.

\subsubsection{ISO Data Archive}

Mid-IR and far-IR spectra of the flow were retrieved from the ISO satellite Data Archive, in the form of Highly Processed Data Product (HPDP set 35500738, by W. Frieswijk et al.; HPDP set 04300333 by Lloyd C., Lerate M., and Grundy T.).
The first set of data (from 2.38 to 45.2\,$\mu$m, see also Tab.~\ref{obs:tab}) was taken with the Short Wavelength Spectrometer (SWS, de Graauw et al.~\cite{degraauw}) in the AOT01 grating mode ($\mathcal R\sim$1000--2000), and is centred on the $IRAS$ source covering a region 
of 33$\arcsec\times$20$\arcsec$.
The covered FoV is shown in Fig.~\ref{slits:fig}.
The second set of data (from 43 to 197\,$\mu$m) was taken with the Long Wavelength Spectrometer (LWS, Clegg et al.~\cite{clegg})
in the AOT01 grating mode ($\lambda/\Delta\lambda$ $\sim$ 200), and is positioned $\sim$48$\arcsec$ north of the source. With a beam of 80$\arcsec$ it covers large parts of the flow, except the so called `jet'.

\section{Results}
\label{results:sec}

\subsection{H$_2$ and [\ion{Fe}{ii}] imaging}
\label{ima_res:sec}
\begin{figure}
 \centering
   \includegraphics [width=7.5 cm] {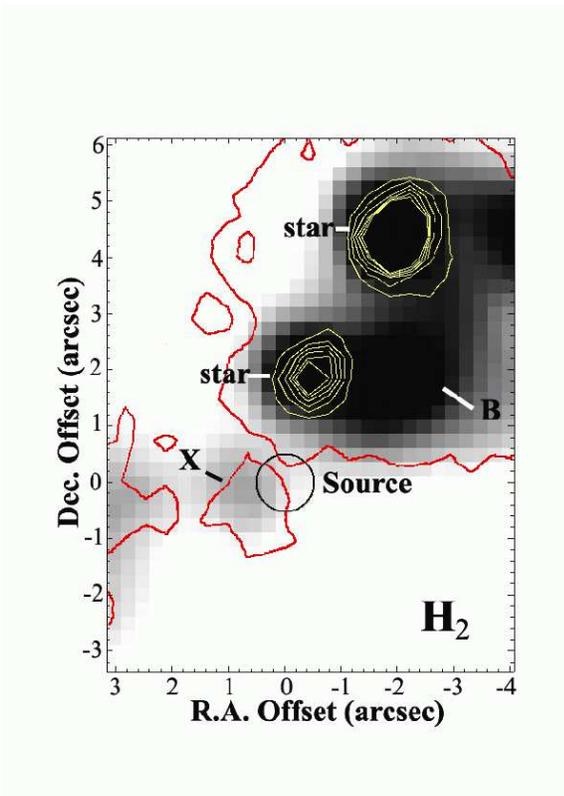}
   \caption{ Close-up of the H$_2$ image around the presumed source position (Hofner et al.~\cite{hofner}, see also the
caption of Fig.~\ref{slits:fig}) indicated by a circle (the radius represents the formal error for the NICS plate solution). [\ion{Fe}{ii}] contours (5 in red, 30 in yellow, 50, 60, 70 ,80 ,100 $\times\sigma\sim$~2.5$\times$~10$^{-17}$~erg~s$^{-1}$~cm$^{-2}$~arcsec$^{-2}$) are overlaid on the image. The positions of the new knot X, knot B, the source, and two field stars are also indicated.
\label{source:fig}}
\end{figure}
\begin{figure}
 \centering
   \includegraphics [width=9.1 cm] {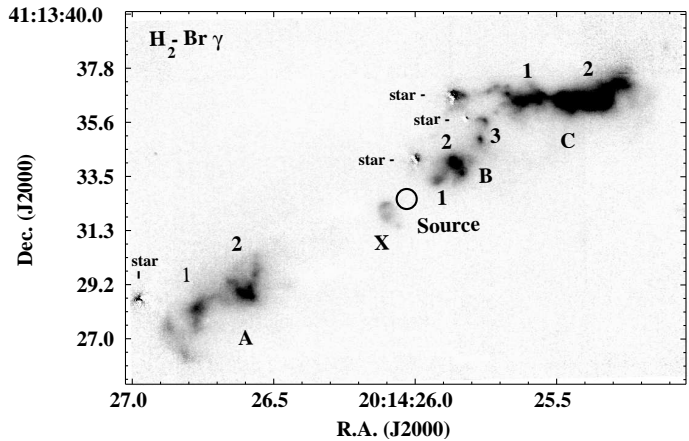}
   \caption{ Continuum-subtracted H$_2$-Br$\gamma$ image from Subaru. 
The Br$\gamma$ image was used as continuum, since no Br$\gamma$ emission is detected on our spectra (see text). The identified knots and the presumed source position (see also the captions of Fig.~\ref{slits:fig} and Fig.~\ref{source:fig}) are indicated. At this scale the wiggling structure of the flow becomes visible.
\label{H2-Brg:fig}}
\end{figure}

An overview of the large-scale morphology of the H$_2$ flow is given in Fig.~\ref{slits:fig}.
The emission in the south-east, the so called `jet', appears to be composed of four different knots, labelled knots 1 to 4
(see Sect.~\ref{highres:sec}). No further knots were detected SE or NW up to $\sim$2\farcm8 (or $\sim$1.4\,pc) from the source, 
down to a 3$\sigma$ limit of $\sim$8$\times$~10$^{-17}$~erg~s$^{-1}$~cm$^{-2}$~arcsec$^{-2}$.

However a new emission feature, labelled X ($\alpha(J2000)$=$20^h14^m26.1^s$ and $\delta(J2000)$=41$\degr$13$\arcmin$32$\arcsec$.4), has been detected close to the position of the $IRAS$ source. Fig.~\ref{source:fig} shows a close-up in H$_2$ of the source position and the new knot, located in the redshifted lobe. The circle represents the formal error for the NICS plate solution centred on 
the IRAS source position (Hofner et al.~\cite{hofner}). 
Contours of the [\ion{Fe}{ii}] emission are superimposed on this image as well. The two stars in the field have been used to match the 
H$_2$ and [\ion{Fe}{ii}] images exactly. Both our H$_2$-K$^\prime$ and [\ion{Fe}{ii}]-H data (as well as the spectroscopic data, see Sect~\ref{lowres:sec}) indicate that the emission of this knot arises from a shock. 
Remarkably, the knot is overlapping with one of the bipolar features (located SE of the source) detected in the K-band by Sridharan et al.~(\cite{sridharan}).
No further [\ion{Fe}{ii}] emission has been detected along this flow, down to a 3$\sigma$ limit of $\sim$3$\times$~10$^{-16}$~erg~s$^{-1}$~cm$^{-2}$~arcsec$^{-2}$, nor along the blue lobe at the position where a H$\alpha$ and [\ion{S}{ii}] emitting knot was observed by Shepherd et al.~(\cite{shepherd}).

The higher angular resolution of Subaru images delineates details of the morphology of the H$_2$ flow close to the source. In Fig.~\ref{H2-Brg:fig} we show the continuum subtracted H$_2$ image. The Br$\gamma$ image turned out more efficient than the K$_{cont}$ image in removing the strong continuum detected on the knots. Its use is justified by the absence of any Br$\gamma$ emission 
in the observed knots (see Sect.~\ref{lowres:sec}). Scattered H$_2$ emission would, of course, not be subtracted
by this procedure.
The H$_2$ flow is wiggling noticeably with an opening angle of $\sim$16$\degr$ around an axis with P.A. of $\sim$-61$\degr$. This fully agrees with SiO radial velocity observations (Cesaroni et al.~\cite{cesaroni99}, Su et al.~\cite{su}), that revealed a precessing jet with an opening cone of 21$\degr$, P.A$\sim$-60$\degr$, and an inclination of 9$\degr$ with respect to the plane of the sky.
This small-amplitude precession seems to be superimposed to the larger one observed at large scale (Shepherd et al.~\cite{shepherd}, 
Cesaroni et al.~\cite{cesaroni05}, see also Fig.~\ref{slits:fig}). 
In the blue lobe (see Fig.~\ref{H2-Brg:fig}) knot B appears composed of a faint jet (feature 1) ending in a bow-shock (2), that is preceeded by two smaller bow-shocks (3). Further ahead, knot C consists of 
at least two bright structures. In the red lobe, the new knot X is visible as well, and, farther out, knot A appears fragmented in 2-3 sub-structures.

\subsection{Low resolution spectroscopy}
\label{lowres:sec}
\subsubsection{Low resolution - TNG}

From our four TNG slits, we have derived low resolution spectra of six knots, namely A, B, C, and X close to the source, and knot 1 and 4 along the `jet'.
All these spectra are rich in H$_2$ emission. Remarkably, no ionic emission is detected along the flow, with the exception 
of faint [\ion{Fe}{ii}] emission in knot X (also observed in the narrow-band image), and in knot B (below the detection limit of
our imaging). It is worth to note that the ratio between the H$_2$ lines (see, e.\,g., 2.12\,$\mu$m) and the 
[\ion{Fe}{ii}] lines changes in the observed knots. Continuum emission from the regions enclosing knots A and B is observed in the K band.

The lines detected in the spectra, along with their vacuum wavelengths, and fluxes (uncorrected for the extinction),
are presented in Table~\ref{spec:tab}.
Line fluxes have been obtained by fitting the profile with a single or double Gaussian in case of 
blending. The uncertainties associated with these data derive only from the rms of the local baseline multiplied by the linewidths. 
Lines showing fluxes with a S/N ratio between 2 and 3 have been labelled in the Table.

\begin{figure*}
 \centering
   \includegraphics [width= 12 cm] {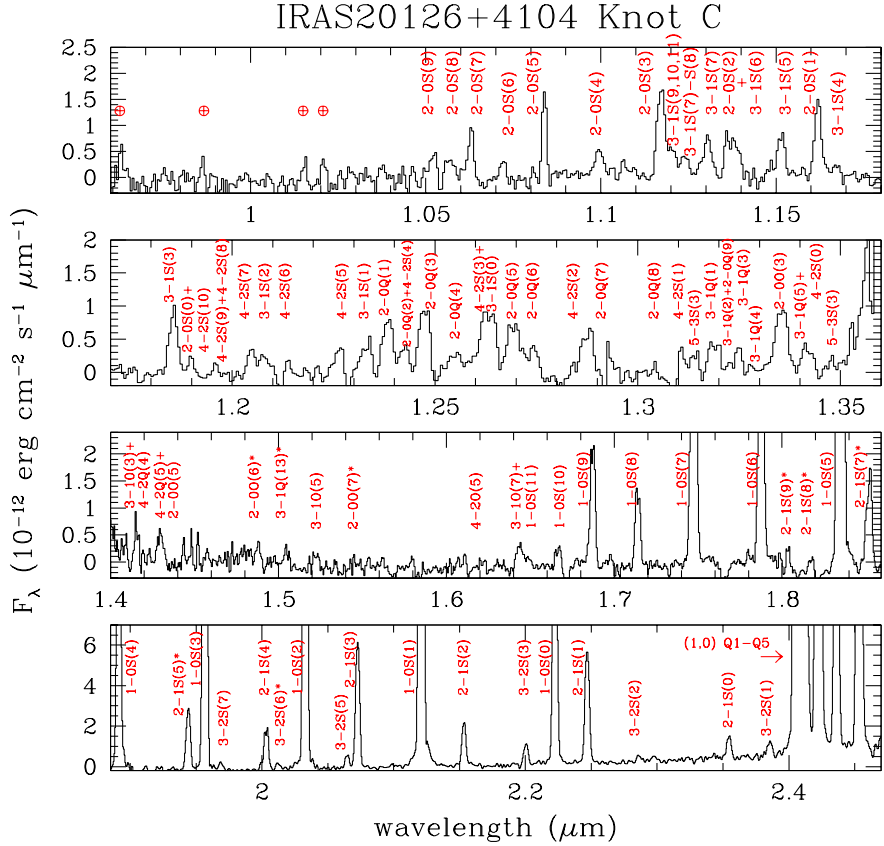}
   \caption{ 0.9--2.5\,$\mu$m low resolution spectrum of knot C in the IRAS\,20126+4104 jet.
   An asterisk near the line identification marks the detections between 2 and 3 sigma. Telluric lines are indicated 
by the symbol ``$\oplus$''.
\label{spectrumC:fig}}
\end{figure*}

Figure~\ref{spectrumC:fig} shows the spectrum of knot C, the brightest
one observed. Only H$_2$ lines are visible from 1 to 2.5\,$\mu$m, involving vibrational levels from v=1 up to v=5
(coming from energy levels from 6000\,K up to 30\,000\,K). The presence of lines from high-v levels (i.\,e. with
high excitation energies) indicates a high excitation of the gas. 
In the remaining knots we identify H$_2$ emission lines up to v$\le$3, mostly detected in the H- and K-bands 
(except for a few v=2 lines from the J-band observed in knot B).

Considering both the combined effects of the knot morphologies and
the slit positions, there is an excellent agreement between the photometric fluxes in our images 
and the fluxes measured in our spectra for both [\ion{Fe}{ii}] (1.64\,$\mu$m) and H$_2$ (2.12\,$\mu$m) lines.

\begin{table*}
\begin{scriptsize}
\caption{Observed lines in the IRAS\,20126+4104 outflow: knots A, B, C, X, jet-1, jet-4. \textit{(To be continued)}\label{spec:tab}}
\begin{center}
\begin{tabular}{cccccccc}
\hline\\[-5pt]
Term&  $\lambda$($\mu$m) & \multicolumn{6}{c}{$F\pm\Delta~F$(10$^{-15}$erg\,cm$^{-2}$\,s$^{-1}$)}\\
\hline\\[-5pt]
 H$_2$ Lines	                &  	 	&    A          &     B    &  C      &   X  & knot 1-jet & knot 4-jet \\
\hline\\[-5pt]
~2--0 S(9)	        & 1.053        & $\cdots$	& $\cdots$	&  1.3$\pm$0.4     & $\cdots$        & $\cdots$        & $\cdots$     \\
~2--0 S(8) 	        & 1.057        & $\cdots$	& $\cdots$	&  0.9$\pm$0.3     & $\cdots$        & $\cdots$        & $\cdots$     \\
~2--0 S(7) 	        & 1.064        & $\cdots$	& $\cdots$	&  2.8$\pm$0.4     & $\cdots$        & $\cdots$        & $\cdots$     \\
~2--0 S(6) 	        & 1.073        & $\cdots$	& $\cdots$	&  1.6$\pm$0.4     & $\cdots$        & $\cdots$        & $\cdots$     \\
~2--0 S(5) 	        & 1.085        & $\cdots$	& $\cdots$	&  3.5$\pm$0.4     & $\cdots$        & $\cdots$        & $\cdots$     \\
~2--0 S(4)	        & 1.100        & $\cdots$	& $\cdots$	&  1.8$\pm$0.3     & $\cdots$        & $\cdots$        & $\cdots$     \\
~2--0 S(3)	        & 1.117        & $\cdots$	& 1.3$\pm$0.4	&  4.3$\pm$0.2     & $\cdots$        & $\cdots$        & $\cdots$     \\
~3--1 S(9)+3--1 S(10)   & 1.120        & $\cdots$	& $\cdots$	&  2.0$\pm$0.2     & $\cdots$        & $\cdots$        & $\cdots$     \\
~3--1 S(11) 	        & 1.122        & $\cdots$	& $\cdots$	&  1.0$\pm$0.3     & $\cdots$        & $\cdots$        & $\cdots$     \\
~3--1 S(8) 	        & 1.125        & $\cdots$	& $\cdots$	&  0.7$\pm$0.2     & $\cdots$        & $\cdots$        & $\cdots$     \\
~3--1 S(7) 	        & 1.130        & $\cdots$	& $\cdots$	&  1.4$\pm$0.2     & $\cdots$        & $\cdots$        & $\cdots$     \\
~3--1 S(13) 	        & 1.132        & $\cdots$	& $\cdots$	&  0.5$\pm$0.2:    & $\cdots$        & $\cdots$        & $\cdots$     \\
~2--0 S(2)  	        & 1.138        & $\cdots$	& $\cdots$	&  2.0$\pm$0.2     & $\cdots$        & $\cdots$        & $\cdots$     \\
~3--1 S(6)    	        & 1.140        & $\cdots$	& $\cdots$	&  0.7$\pm$0.2     & $\cdots$        & $\cdots$        & $\cdots$     \\
~3--1 S(5)	        & 1.152        & $\cdots$	& $\cdots$	&  1.9$\pm$0.3     & $\cdots$        & $\cdots$        & $\cdots$     \\
~2--0 S(1) 	        & 1.162        & $\cdots$	& 1.3$\pm$0.4	&  3.5$\pm$0.3     & $\cdots$        & $\cdots$        & $\cdots$     \\
~3--1 S(4)	        & 1.167        & $\cdots$	& $\cdots$	&  1.2$\pm$0.3     & $\cdots$        & $\cdots$        & $\cdots$     \\
~3--1 S(3)	        & 1.186        & $\cdots$	& $\cdots$	&  2.5$\pm$0.3     & $\cdots$        & $\cdots$        & $\cdots$     \\
~2--0 S(0)+4--2 S(10)   & 1.189-1.190  & $\cdots$	& $\cdots$	&  0.9$\pm$0.2     & $\cdots$        & $\cdots$        & $\cdots$     \\
~4--2 S(9)+4--2 S(8)    & 1.196-1.199  & $\cdots$	& $\cdots$	&  0.7$\pm$0.2     & $\cdots$        & $\cdots$        & $\cdots$      \\
~4--2 S(7)	        & 1.205        & $\cdots$	& $\cdots$	&  1.1$\pm$0.2     & $\cdots$        & $\cdots$        & $\cdots$     \\
~3--1 S(2)	        & 1.207        & $\cdots$	& $\cdots$	&  0.9$\pm$0.2     & $\cdots$        & $\cdots$        & $\cdots$     \\
~4--2 S(6) 	        & 1.214        & $\cdots$	& $\cdots$	&  0.6$\pm$0.2     & $\cdots$        & $\cdots$        & $\cdots$     \\
~4--2 S(5) 	        & 1.226        & $\cdots$	& $\cdots$	&  1.4$\pm$0.2     & $\cdots$        & $\cdots$        & $\cdots$     \\
~3--1 S(1) 	        & 1.233        & $\cdots$	& $\cdots$	&  1.4$\pm$0.2     & $\cdots$        & $\cdots$        & $\cdots$     \\
~2--0 Q(1) 	        & 1.238        & $\cdots$	&0.6$\pm$0.3:	&  2.6$\pm$0.2     & $\cdots$        & $\cdots$        & $\cdots$     \\
~2--0 Q(2)+4--2 S(4)    & 1.242-1.242  & $\cdots$	& $\cdots$	&  1.4$\pm$0.2     & $\cdots$        & $\cdots$        & $\cdots$     \\
~2--0 Q(3) 	        & 1.247        & $\cdots$	&0.9$\pm$0.4:	&  2.8$\pm$0.2     & $\cdots$        & $\cdots$        & $\cdots$     \\
~2--0 Q(4) 	        & 1.254        & $\cdots$	& $\cdots$	&  1.1$\pm$0.2     & $\cdots$        & $\cdots$        & $\cdots$     \\
~4--2 S(3)+3--1 S(0)    & 1.261        & $\cdots$	& $\cdots$	&  2.7$\pm$0.3     & $\cdots$        & $\cdots$        & $\cdots$     \\
~2--0 Q(5)              & 1.263        & $\cdots$	& $\cdots$	&  2.1$\pm$0.3     & $\cdots$        & $\cdots$        & $\cdots$     \\
~2--0 Q(6)	        & 1.274        & $\cdots$	& $\cdots$	&  0.7$\pm$0.2     & $\cdots$        & $\cdots$        & $\cdots$     \\
~4--2 S(2)              & 1.284        & $\cdots$	& $\cdots$	&  0.6$\pm$0.2     & $\cdots$        & $\cdots$        & $\cdots$     \\
~2--0 Q(7)	        & 1.287        & $\cdots$	& $\cdots$	&  1.9$\pm$0.2     & $\cdots$        & $\cdots$        & $\cdots$     \\
~2--0 Q(8)	        & 1.302        & $\cdots$	& $\cdots$	&  0.7$\pm$0.2     & $\cdots$        & $\cdots$        & $\cdots$     \\
~4--2 S(1)	        & 1.311        & $\cdots$	& $\cdots$	&  0.9$\pm$0.2     & $\cdots$        & $\cdots$        & $\cdots$     \\
~5--3 S(5)	        & 1.312        & $\cdots$	& $\cdots$	&  0.7$\pm$0.2     & $\cdots$        & $\cdots$        & $\cdots$     \\
~3--1 Q(1)              & 1.314        & $\cdots$	& $\cdots$	&  1.2$\pm$0.2     & $\cdots$        & $\cdots$        & $\cdots$     \\
~3--1 Q(2)+2--0 Q(9)    & 1.318-1.319  & $\cdots$	& $\cdots$	&  1.6$\pm$0.2     & $\cdots$        & $\cdots$        & $\cdots$     \\
~3--1 Q(3) 	        & 1.324        & $\cdots$	& $\cdots$	&  1.3$\pm$0.2     & $\cdots$        & $\cdots$        & $\cdots$     \\
~3--1 Q(4)  	        & 1.333        & $\cdots$	& $\cdots$	&  0.5$\pm$0.2:    & $\cdots$        & $\cdots$        & $\cdots$     \\
~2--0 O(3) 	        & 1.335        & $\cdots$	& $\cdots$	&  2.7$\pm$0.2     & $\cdots$        & $\cdots$        & $\cdots$     \\
~3--1 Q(5)+4--2 S(0)    & 1.342-1.342  & $\cdots$	& $\cdots$	&  1.6$\pm$0.2     & $\cdots$        & $\cdots$        & $\cdots$     \\
~5--3 S(3) 	        & 1.347        & $\cdots$	& $\cdots$	&  0.7$\pm$0.2     & $\cdots$        & $\cdots$        & $\cdots$     \\
~3--1 O(3)+4--2 Q(4)    & 1.418        & $\cdots$	& $\cdots$	&  1.3$\pm$0.4     & $\cdots$        & $\cdots$        & $\cdots$     \\
~4--1 Q(5)+2-0 O(5)     & 1.430-1.432  & $\cdots$	& $\cdots$	&  2.1$\pm$0.4     & $\cdots$        & $\cdots$        & $\cdots$     \\
\hline
\end{tabular}
\end{center}
\end{scriptsize}
\end{table*}
\addtocounter{table}{-1}
\begin{table*}
\begin{scriptsize}
\caption{Observed lines in the IRAS\,20126+4104 outflow: knots A, B, C, X, jet-1, jet-4. \textit{(Continued)}}
\begin{center}
\begin{tabular}{cccccccc}
\hline\\[-5pt]
Term&  $\lambda$($\mu$m) & \multicolumn{6}{c}{$F\pm\Delta~F$(10$^{-15}$erg\,cm$^{-2}$\,s$^{-1}$)}\\
\hline\\[-5pt]
 H$_2$ Lines	                &  	 	&    A          &     B    &  C      &   X  & knot 1-jet & knot 4-jet \\
\hline\\[-5pt]
~2--0 O(6)              & 1.487        & $\cdots$	& $\cdots$	&  1.0$\pm$0.4:    & $\cdots$        & $\cdots$        & $\cdots$     \\
~3--1 Q(13)             & 1.502        & $\cdots$	& $\cdots$	&  0.6$\pm$0.3:    & $\cdots$        & $\cdots$        & $\cdots$     \\
~3--1 O(5)	        & 1.522        & $\cdots$	& $\cdots$	&  1.3$\pm$0.4     & $\cdots$        & $\cdots$        & $\cdots$     \\
~2--0 O(7) 	        & 1.545        & $\cdots$	& $\cdots$	&  0.7$\pm$0.3:    & $\cdots$        & $\cdots$        & $\cdots$     \\
~4--2 O(5)	        & 1.622        & $\cdots$	& $\cdots$	&  0.9$\pm$0.3     & $\cdots$        & $\cdots$        & $\cdots$     \\
~3--1 O(7) 	        & 1.645        & $\cdots$	& $\cdots$	&  0.7$\pm$0.2     & $\cdots$        & $\cdots$        & $\cdots$     \\
~1--0 S(11) 	        & 1.650        & $\cdots$	& $\cdots$	&  1.0$\pm$0.3     & $\cdots$        & $\cdots$        & $\cdots$     \\
~1--0 S(10) 	        & 1.666        & $\cdots$	& $\cdots$	&  1.1$\pm$0.3     & $\cdots$        & $\cdots$        & $\cdots$     \\
~1--0 S(9) 	        & 1.688        & 1.9$\pm$0.4	& 2.1$\pm$0.3	&  7.4$\pm$0.3     & $\cdots$      & 0.9$\pm$0.3       & 0.7$\pm$0.3:	 \\
~1--0 S(8) 	        & 1.715        & 1.4$\pm$0.4	& 1.2$\pm$0.2	&  5.2$\pm$0.3     & $\cdots$      & 0.7$\pm$0.3:      & 0.7$\pm$0.3:  	 \\
~1--0 S(7) 	        & 1.748        & 12.1$\pm$0.4	& 7.5$\pm$0.2	& 33.7$\pm$0.3   & 1.9$\pm$0.2     & 3.8$\pm$0.3       & 2.5$\pm$0.3 	  \\
~1--0 S(6) 	        & 1.788        & 7.9$\pm$0.4	& 5.0$\pm$0.2	& 22.8$\pm$0.3   & 1.4$\pm$0.2     & 2.4$\pm$0.3       & 1.5$\pm$0.4 	  \\
~2--1 S(9) 	        & 1.790        & $\cdots$	& $\cdots$	&  1.0$\pm$0.5:    & $\cdots$        & $\cdots$        & $\cdots$     \\

~2--1 S(8) 	        & 1.818        & $\cdots$	& $\cdots$	&  1.0$\pm$0.5:    & $\cdots$        & $\cdots$        & $\cdots$     \\
~1--0 S(5) 	        & 1.836        & 54$\pm$5	& 31$\pm$5	& 125$\pm$5	   & $\cdots$        & $\cdots$        & $\cdots$     \\
~2--1 S(7) 	        & 1.853        & $\cdots$	& $\cdots$	&  8.0$\pm$5:      & $\cdots$        & $\cdots$        & $\cdots$     \\
~1--0 S(4) 		 & 1.891        & 28$\pm$5   	 & 16$\pm$5:	 &  66$\pm$5	    & $\cdots$  & $\cdots$        & $\cdots$  \\
~2--1 S(7) 		 & 1.945        & $\cdots$       & $\cdots$      & 12.0$\pm$5: & $\cdots$  & $\cdots$        & $\cdots$  \\
~1--0 S(3) 		 & 1.958        & 24$\pm$2   	 & 51$\pm$5  	 & 110$\pm$5	    & 13$\pm$4  & $\cdots$        & $\cdots$  \\
~3--2 S(7)           	& 1.969        & $\cdots$	 & $\cdots$	 & 1.7$\pm$0.8:  & $\cdots$ & $\cdots$        & $\cdots$   \\
~2--1 S(4)           	& 2.004     &  $\cdots$ 	 &  1.3$\pm$0.4       &  8.3$\pm$0.8      &  $\cdots$	   & 0.6$\pm$0.3:    & $\cdots$ \\
~3--2 S(6)           	& 2.013     &  $\cdots$ 	 &  $\cdots$	      &  1.2$\pm$0.5:     &  $\cdots$	   & $\cdots$	     & $\cdots$   \\
~1--0 S(2)           	& 2.034     &  38.9$\pm$0.5	 &  21.1$\pm$0.4      &  83.4$\pm$0.5     &  4.2$\pm$0.4   & 7.7$\pm$0.3     &  5.7$\pm$0.4  \\
~3--2 S(5)           	& 2.066     &  0.9$\pm$0.3	 &  0.7$\pm$0.3:      &  2.9$\pm$0.4      &  $\cdots$	   & $\cdots$	     & $\cdots$   \\
~2--1 S(3)           	& 2.073     &  7.4$\pm$0.4	 &  4.8$\pm$0.3       &  24.4$\pm$0.4     &  1.4$\pm$0.4   & 2.0$\pm$0.4     & 1.1$\pm$0.3    \\
~1--0 S(1)           	& 2.122     & 121$\pm$0.4	 & 66.4$\pm$0.4       & 246$\pm$0.4       & 11.1$\pm$0.4   & 22.6$\pm$0.4    & 16.6$\pm$0.4    \\
~2--1 S(2)           	& 2.154     &  3.7$\pm$0.4	 &  2.2$\pm$0.4       &  8.4$\pm$0.4      &  0.5$\pm$0.2:  &  1.0$\pm$0.3    &  0.6$\pm$0.2    \\
~3--2 S(3)           	& 2.201     &  1.6$\pm$0.4	 &  0.8$\pm$0.3:      &  4.3$\pm$0.4      &  0.4$\pm$0.1:  &  0.5$\pm$0.2:   &   $\cdots$      \\
~1--0 S(0)           	& 2.223     &  30.1$\pm$0.4	 &  15.6$\pm$0.4      &  59.5$\pm$0.4     &  2.8$\pm$0.3   &  5.7$\pm$0.3    &  4.6$\pm$0.4  \\
~2--1 S(1)           	& 2.248     &  10.9$\pm$0.4	 &  5.5$\pm$0.4       &  23.0$\pm$0.4     &  1.6$\pm$0.4   &  2.5$\pm$0.3    &  2.1$\pm$0.4  \\
~3--2 S(2)           	& 2.286     &  0.9$\pm$0.4:	&  $\cdots$	      &  1.5$\pm$0.4      &  $\cdots$	   & $\cdots$	     & $\cdots$   \\
~2--1 S(0)           	& 2.355     &  2.7$\pm$0.5	&  1.7$\pm$0.5	      &  5.1$\pm$0.5      &  $\cdots$	   & $\cdots$	     & $\cdots$    \\
~3--2 S(1)           	& 2.386     &  1.0$\pm$0.5:	&  $\cdots$	      &  4.0$\pm$0.5      &  $\cdots$	   & $\cdots$	     & $\cdots$   \\
~1--0 Q(1)           	& 2.407     & 135$\pm$5 	&  66$\pm$5           & 234$\pm$5         &  11$\pm$3	   & 21$\pm$5	     &  16$\pm$5      \\
~1--0 Q(2)           	& 2.413     &  50$\pm$5 	&  25$\pm$5	      &  89$\pm$5         &  $\cdots$	   & $\cdots$	     & $\cdots$   \\
~1--0 Q(3)           	& 2.424     & 123$\pm$5 	&  57$\pm$5	      & 218$\pm$5         & 11$\pm$3	   & 20$\pm$5	     &  13$\pm$5:   \\
~1--0 Q(4)           	& 2.437     &  35$\pm$5 	&  18$\pm$5	      &  70$\pm$5         &  $\cdots$	   & $\cdots$	     & $\cdots$   \\
~1--0 Q(5)           	& 2.455     &  39$\pm$5 	&  27$\pm$5	      & 142$\pm$5         &  $\cdots$	   & $\cdots$	     & $\cdots$   \\
\hline\\[-5pt]
  [{\ion{Fe}{II}}] lines         &              &                &                &               &     \\
\hline\\[-5pt]
~$a^4\!D_{7/2}-a^6\!D_{9/2}$ & 1.257 	& $\cdots$   	& 0.6$\pm$0.3$^a$ &   $\cdots$  &    1.2$\pm$0.3	& $\cdots$	     & $\cdots$\\
~$a^4\!D_{7/2}-a^6\!D_{7/2}$ & 1.321 	& $\cdots$  	& $\cdots$        &   $\cdots$  &    0.6$\pm$0.3$^a$	& $\cdots$	     & $\cdots$\\
~$a^4\!D_{7/2}-a^4\!F_{9/2}$ & 1.644 	& $\cdots$  	& 1.7$\pm$0.2     &   $\cdots$  &    2.0$\pm$0.2	& $\cdots$	     & $\cdots$\\
~$a^4\!D_{7/2}-a^4\!F_{7/2}$+$a^4\!P_{5/2}-a^4\!D_{7/2}$  & 1.810-1.811 & $\cdots$   & 1.2$\pm$0.4       &   $\cdots$    & 1.5$\pm$0.4	& $\cdots$	     & $\cdots$\\            & \\
\hline\\[-5pt]
\hline
\end{tabular}
\end{center}
Notes: 3$\sigma$ upper limits (in knot C) for the [\ion{C}{i}] doublet (at 0.983-5\,$\mu$m) and for
the Br$\gamma$ line are 10$^{-15}$ and 6$\times$10$^{-16}$\,erg\,cm$^{-2}$\,s$^{-1}$, respectively.
\end{scriptsize}
\end{table*}

\begin{table}
\begin{scriptsize}
\begin{center}
\caption{Observed lines in the IRAS\,20126+4104 outflow: ISO lines }
\label{iso:tab}
\begin{tabular}{ccc}
\hline\\[-5pt]
Term&  $\lambda$($\mu$m) & $F\pm\Delta~F$(10$^{-13}$erg\,cm$^{-2}$\,s$^{-1}$)\\
\hline\\[-5pt]
 H$_2$ Lines	                &  	 	&    Flux             \\
\hline\\[-5pt]
~1--0 Q(1) 	     & 2.407	    & 1.8$\pm$0.4 	           \\
~1--0 Q(2) 	     & 2.413	    & 1.4$\pm$0.4 	           \\
~1--0 Q(3) 	     & 2.424	    & 0.8$\pm$0.4 	           \\
~1--0 Q(4) 	     & 2.437	    & 0.7$\pm$0.3: 	           \\
~1--0 Q(5) 	     & 2.454	    & 1.4$\pm$0.3 	           \\
~1--0 Q(6) 	     & 2.476	    & 1.3$\pm$0.2 	           \\
~1--0 Q(7) 	     &  2.500	    & 1.4$\pm$0.1 	           \\
~1--0 O(2) 	     &  2.627	    & 0.6$\pm$0.1 	           \\
~1--0 O(3) 	     &  2.803	    & 0.7$\pm$0.1 	           \\
~1--0 O(4) 	     &  3.004	    & 0.6$\pm$0.2 	           \\
~1--0 O(5) 	     &  3.235	    & 0.8$\pm$0.1 	           \\
~0--0 S(6) 	     &  6.109	    & 21$\pm$3      \\
~0--0 S(5) 	     &  6.909	    & 21$\pm$3      \\
~0--0 S(4) 	     &  8.026	    & 30$\pm$3       \\
~0--0 S(3) 	     &  9.665	    & 32$\pm$6      \\
~0--0 S(2) 	     & 12.278	    & 26$\pm$8       \\
~0--0 S(1) 	     & 17.034	    & 56$\pm$8       \\
\hline\\[-5pt]
  ionic lines                &               &     \\
\hline\\[-5pt]
~[\ion{O}{i}]$a^3\!P_{1}-a^3\!P_{2}$ & 63.18 	& 260$\pm$10	\\
~[\ion{C}{ii}]$a^3\!P_{3/2}-a^3\!P_{1/2}$ & 157.74 	& 370$\pm$10	\\
\hline\\[-5pt]
\hline
\end{tabular}
\end{center}
\end{scriptsize}
\end{table}

\subsubsection{Low resolution - ISO}
\label{lowresISO:sec}

\begin{figure}
 \centering
   \includegraphics [width= 9.0 cm] {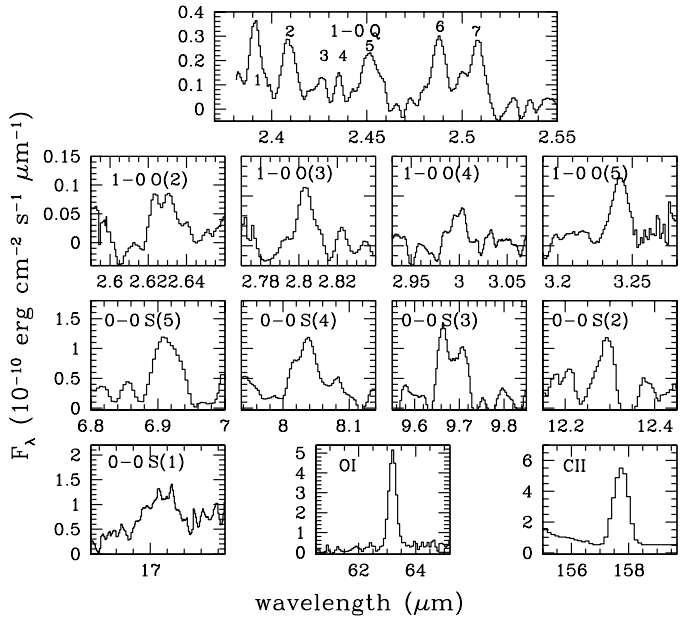}
   \caption{ ISO- SWS and LWS detected lines.
\label{spectrumISO:fig}}
\end{figure}

Table~\ref{iso:tab} lists the lines identified in the ISO-SWS and ISO-LWS spectra, their vacuum wavelengths and integrated fluxes, 
estimated from Gaussian fits to the (unresolved) line profiles, together with the error from the rms noise of the local baseline.
Due to the large FoV of the two instruments, it is not possible to determine the spatial extension of the emitting region, thus 
the measured fluxes can originate from any or all of the different knots (A$\cdots$D) close to the source.
Several emssion lines are detected, superimposed on a strong continuum observed in both SWS and LWS spectra.
The detected lines are shown in Fig.~\ref{spectrumISO:fig}.
H$_2$ lines from the v=1 and v=0 levels have been detected between 2.4 and 17\,$\mu$m. 
In particular, the 1-0\,Q lines (also observed in our NIR spectra), the 1-0\,O lines, and the 0-0\,S lines, 
that usually trace the coldest component of the H$_2$ flow (see e.\,g., Froebrich et al.~\cite{froebrich02}, Giannini et al.~\cite{giannini04}, Giannini et al.~\cite{giannini06}) have been detected. In addition,
two bright ionic lines have been observed in the LWS spectrum, namely [\ion{O}{i}] at 63\,$\mu$m and [\ion{C}{ii}]
at 158\,$\mu$m. The origin of these lines is indeed intriguing, because no strong ionic emission lines were detected in the NIR. Possibly, they could originate from the same spot where [\ion{Fe}{ii}] is detected, and/or in a deeply embedded region 
(close to the source) not visible at NIR wavelengths, because our spectra 
do not show any evidence of the bright [\ion{C}{i}] doublet at 1\,$\mu$m, often detected in low-mass jets 
(see e.\,g. Nisini et al.~\cite{nisini}, Giannini et al.~\cite{giannini04}).
However, a more convincing explanation is that these lines are not arising from the shocks, but from a \textit{photo-dissociation region}
(PDR) around the massive object. Indeed, from the [\ion{O}{i}](63\,$\mu$m)/[\ion{C}{ii}](158\,$\mu$m) ratio, we can discriminate between the two scenarios (see e.\,g., Hollenbach \& McKee~\cite{hollen}). For intensity ratios below 10 the PDR origin is favoured. In our LWS spectrum this ratio is close to 1.

\subsection{High resolution spectroscopy}
\label{highres:sec}

\begin{figure*}
 \centering
   \includegraphics [width=16 cm] {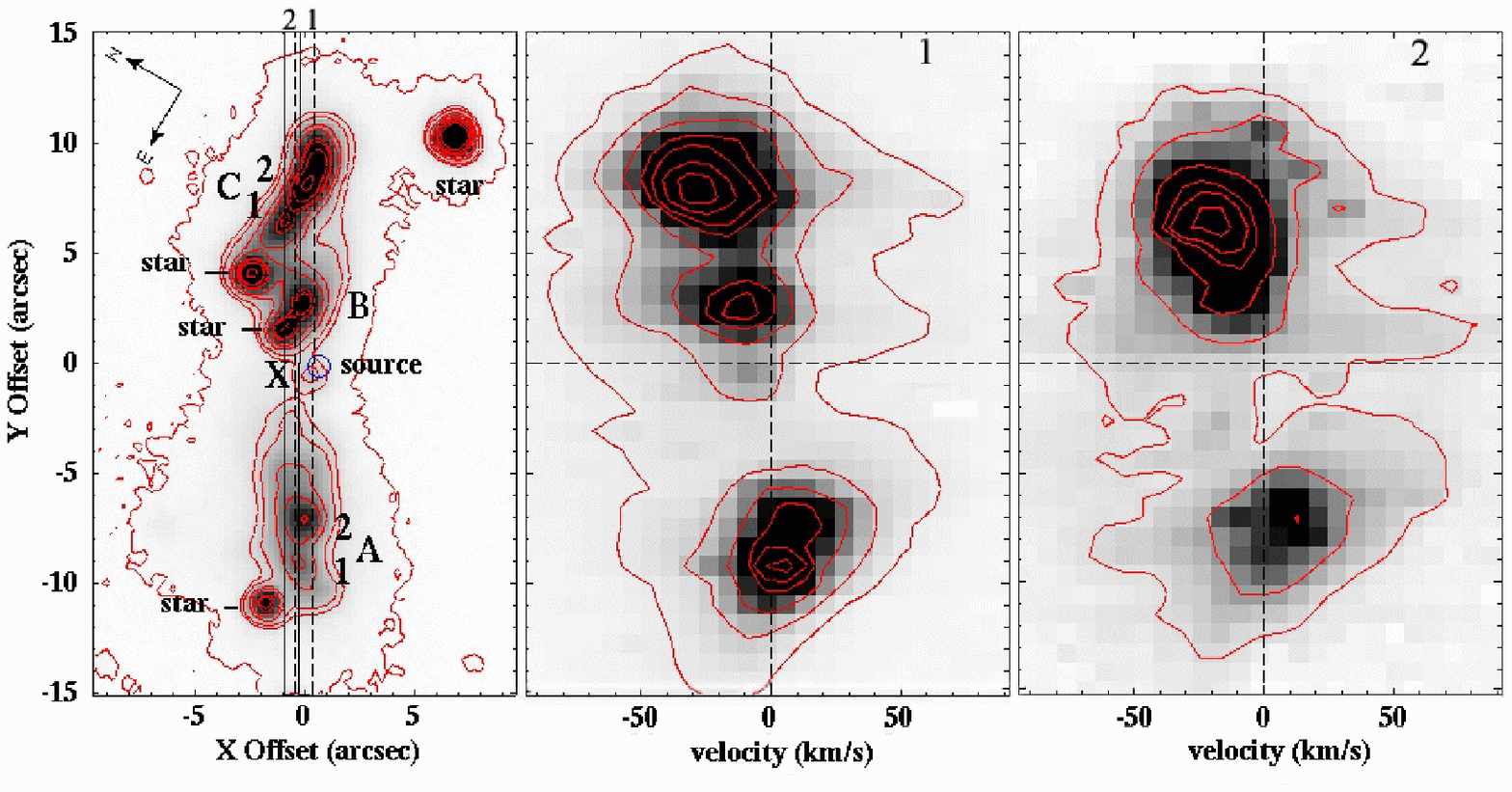}
   \caption{ P-V diagram for slits 1 and 2. The {\bf left} panel shows the H$_2$ (TNG) image with the slits 
superimposed, and the position of the knots. Y-offsets are displayed from the position of the source.
The contour levels are 3, 30, 50, 100, 130, 210, 235, 300, 420$\times$ the standard deviation to the mean background ($\sigma\sim$~2~$\times$~10$^{-17}$~erg~s$^{-1}$~cm$^{-2}$~arcsec$^{-2}$).
The {\bf central} and {\bf right} panels display the uncalibrated H$_2$ (2.12\,$\mu$m) spectra from CGS4
(slits 1 and 2, respectively). 
The contour levels are 5, 20, 60, 125, 180, 300, 450$\times$ the standard deviation to the mean background.
\label{PV1:fig}}
\end{figure*}

\begin{figure*}
 \centering
   \includegraphics [width=9.0 cm] {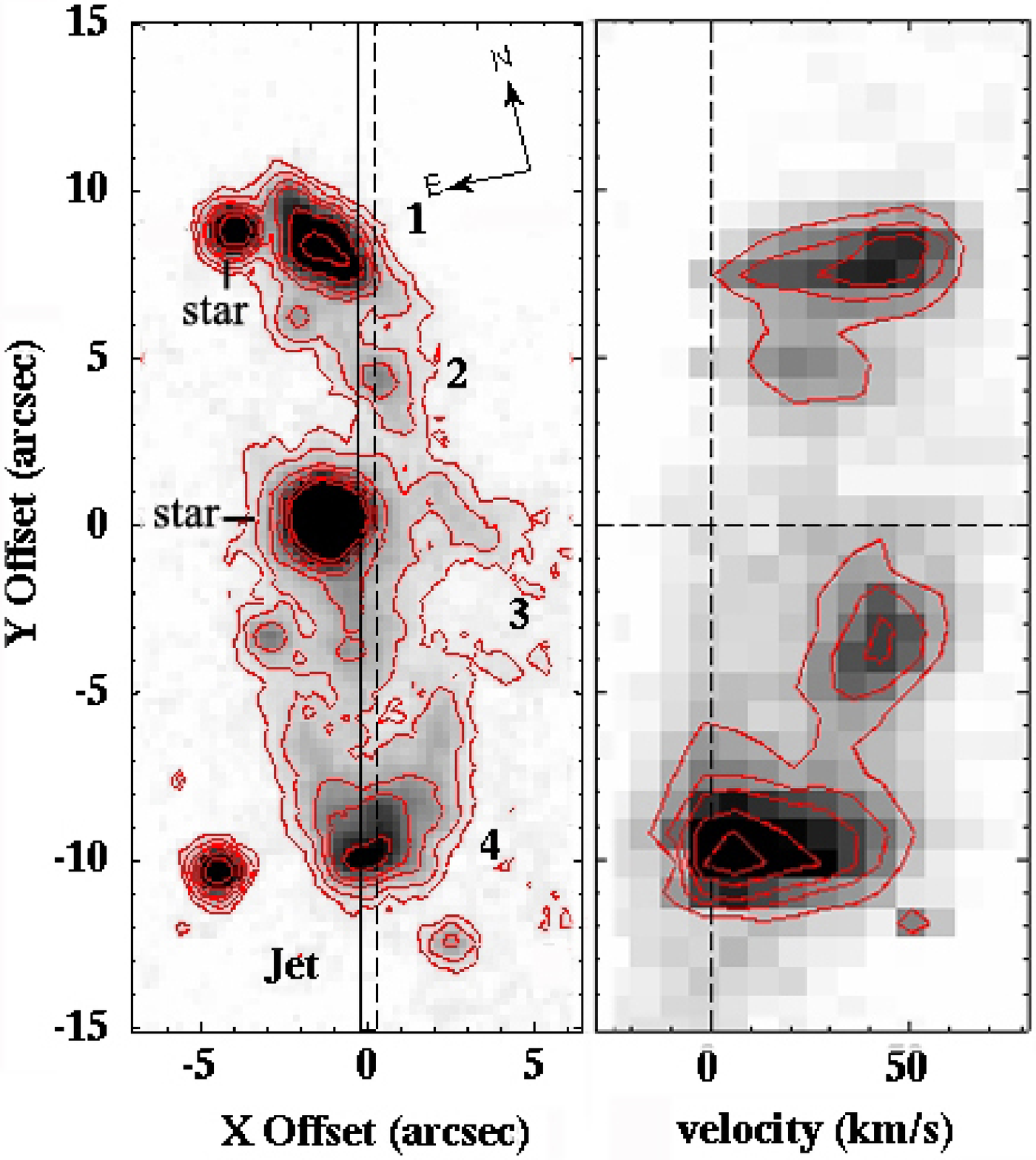}
\includegraphics [width=9.0 cm] {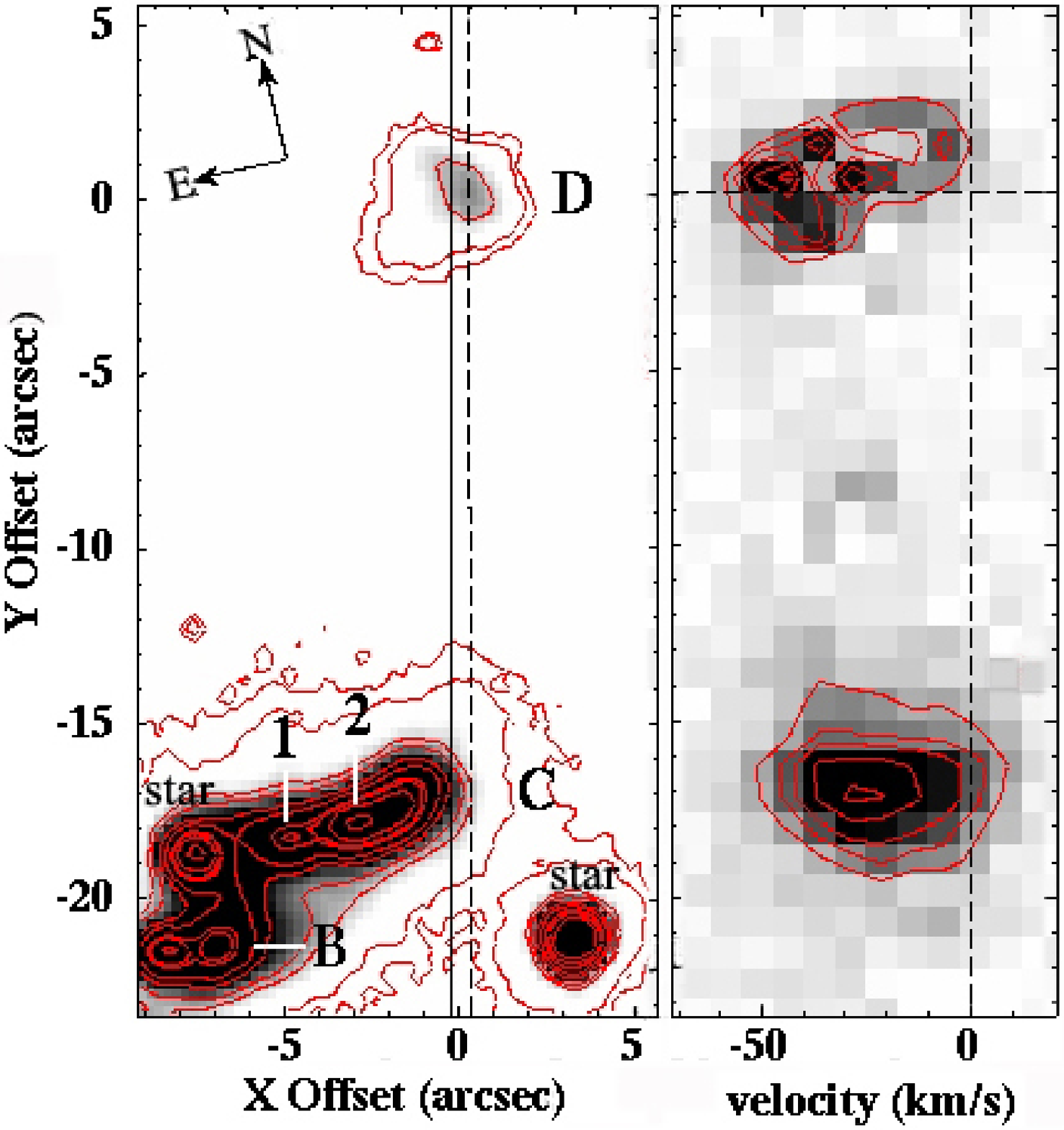}
   \caption{\textbf{Left} P-V diagram for slit 3. On the {\bf left} panel the H$_2$ (TNG) image indicates the position of the slit and knots. Y offsets are displayed from the position of the star in the centre of the image. The contour levels are 3, 6, 12, 20, 30, 50$\times$~$\sigma$. On the {\bf right} the uncalibrated H$_2$ (2.12\,$\mu$m) spectrum from CGS4 is reported. The contour levels are 3, 5, 7, 10, 13$\times$~$\sigma$.
\textbf{Right} P-V diagram for slit 4. Y offsets are displayed from the position of knot D. The image contour levels are 3, 6, 30, 50, 100, 130, 210, 235, 300, 420$\times$~$\sigma$. The levels of the spectra are 3, 5, 7, 10, 13$\times$~$\sigma$.
\label{PV2:fig}}
\end{figure*}

\begin{table}
\caption{ H$_2$ radial velocities of individual knots in the IRAS\,20126+4104 flow from CGS4 observations.  \label{H2vrad:tab}}
\vspace {0.5 cm}
\begin{center}
\begin{tabular}{ccc}
\multicolumn{3}{c}{CGS4 slit 1}\\ \hline \hline
knot  & $v_{rad}$       & FWZI  \\
      & (km\,s$^{-1}$)   & (km\,s$^{-1}$)   \\
\hline
A1    & 2$\pm$3 & 110   \\
A2    & 9$\pm$3 & 150   \\
X     & -8$\pm$3; -28$\pm$3; 53$\pm$3   & 175  \\
B     & -14$\pm$3  & 160   \\
C2    & -25$\pm$3  & 160   \\
\hline\\[-5pt]
\multicolumn{3}{c}{CGS4 slit 2}\\ \hline \hline
A1    &  -1$\pm$3; -52$\pm$3    & 130  \\
A2    &  9$\pm$3; -12$\pm$3; 53$\pm$3  & 180   \\
B     & -14$\pm$3; -46$\pm$3; 26$\pm$3; 64$\pm$3  & 170   \\
C1    & -19$\pm$3  & 130   \\
\hline\\[-5pt]
\multicolumn{3}{c}{CGS4 slit 3}\\ \hline \hline
jet knot\,1    &  47$\pm$3; 19$\pm$3    & 70   \\
jet knot\,2    &  27$\pm$3   & 75   \\
jet knot\,3    &  41$\pm$3;  2$\pm$3   & 90   \\
jet knot\,4    &   6$\pm$3;  36$\pm$3  & 90   \\
\hline\\[-5pt]
\multicolumn{3}{c}{CGS4 slit 4}\\ \hline \hline
D     & -42$\pm$3; -17$\pm$3  & 70   \\
C2    & -28$\pm$3; -5$\pm$3; 3$\pm$3  & 90   \\
\hline\hline
\end{tabular}
\end{center}
\end{table}

In Table~\ref{H2vrad:tab} we report the results of the high resolution H$_2$ spectroscopy. 
For each knot in each slit  we indicate the H$_2$ radial velocities ($v_{rad}$), corrected for the cloud speed with respect to the LSR ($v_{LSR}=-3.5$\,km\,s$^{-1}$, Cesaroni et al.~\cite{cesaroni97}). Two or more velocity components are given
where detected. The first reported value is the peak value.
The last column gives the {\it full width at zero intensity (FWZI)} of the line profile,
measured where the flux reaches a 2$\sigma$ background noise level (see e.\,g. Davis et al.~\cite{davis01}).

\begin{figure}
 \centering
   \includegraphics [width=9.1 cm] {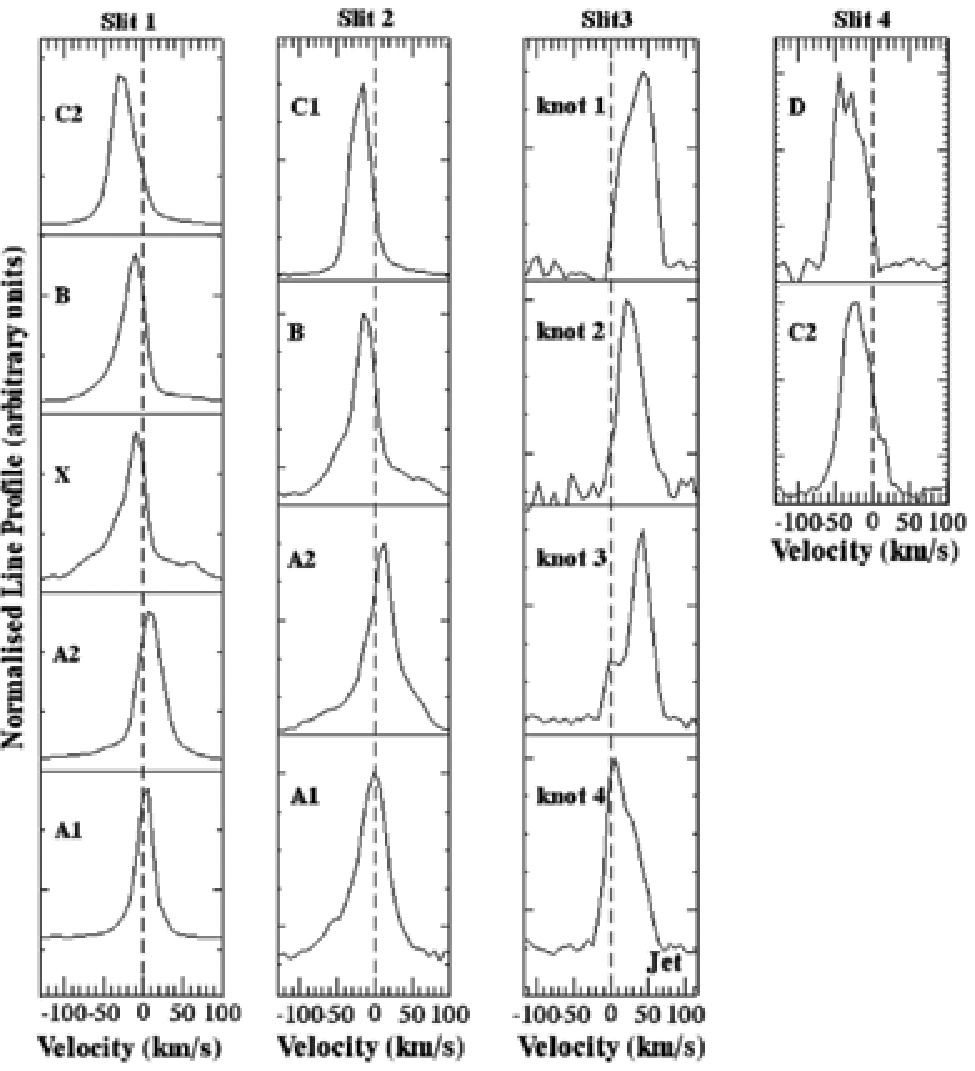}
   \caption{ Line profiles of the knots in the 1-0\,S(1) line of H$_2$ at 2.12\,$\mu$m.
\label{profile:fig}}
\end{figure}

Position-velocity (P-V) diagrams are presented in Figure~\ref{PV1:fig} (CGS4 Slits 1 and 2), and in Fig.~\ref{PV2:fig} (CGS4 Slits 3 and 4). 
In Fig.~\ref{profile:fig} the emission line profiles are shown.
It is worth to note that our spectra often cannot resolve the single structures inside
the wiggling jet, due to the limited spatial resolution. As a result, radial velocities are often an average over more than one sub-structure visible in the Subaru image.

Almost all the knots also have more than one velocity component in their spectral line profile, depending on the position 
of the slit with respect to the knot.
However, such components are never completely resolved (with the exception of knot 2 along the `jet', see Fig.~\ref{profile:fig}),
but they rather show up like `bumps' along the smooth profile.
When single-peaked profiles are observed, the lines are never symmetric, but evidence for extended line-wing emission (opposite
to the blue- or red-shifted peak) is always detected. This may be indicative of a bow-shock morphology of the knots, producing
both the wings and the different velocity components (see e.\,g. Davis et al.~\cite{davis01}, Schultz et al.~\cite{schultz}),
as also indicated by the sub-millimetric observations of Su et al.~(\cite{su}).
Remarkably, knot B has four velocity components (see Tab.~\ref{H2vrad:tab}, and Fig.~\ref{profile:fig}). 
The fourth component located at $\sim$64\,km\,s$^{-1}$, and visible along the profile as a `bump' on the red-shifted wing (see also knot X profile), cannot be explained in a bow-shock context. This component could originate from a different flow, but we have no further 
evidence in support of this hypothesis.

The flux peaks of the knots range from -42 to -14\,km\,s$^{-1}$ in the blue lobe, and from -8 to 47\,km\,s$^{-1}$ in the red lobe (see also Tab.~\ref{H2vrad:tab}). Noticeably, knot X and A in the red lobe (Fig.~\ref{PV1:fig}, central panel) have a slightly negative peak velocity. This can be understood considering that close to the source the axis of the flow has an inclination of $\sim$9$\degr$ with respect to the plane of the sky, and the aperture angle of the precessing jet is $\sim$37$\degr$. Consequently, even if the knots are located in the `red' lobe, they could have a negative $v_{rad}$ (see also Su et al.~\cite{su}).

In both lobes, the absolute peak radial velocities of knots close to the source (A$\cdots$C) are smaller (0--30\,km\,s$^{-1}$) than those located at larger distances (knot D and knots 1$\cdots$4, 40--50\,km\,s$^{-1}$).
On the other hand, the \textit{FWZI} of the line profiles decreases with distance (see Tab.~\ref{H2vrad:tab} and Fig.~\ref{profile:fig}). 
The line profiles of the knots close to the source are broader on average (110--180\,km\,s$^{-1}$) than those far from the source (70--90\,km\,s$^{-1}$). This could confirm that the inclination of the flow axis (with respect to the plane of the sky) is different in the two regions (i.\,e. it changes from $\sim$9$\degr$, close to the source, to $\sim$45$\degr$ further out, at knots 1$\cdots$4 and knot D).
On smaller scale, radial velocities appear to oscillate. This is well visible in P-V diagrams of Slit 3, and marginally,
for the red-shifted knots in Slit 1.

Assuming two different values for the inclination, the spatial velocity of the knots ranges between 50 and 80\,km\,s$^{-1}$.

\subsection{Physical parameters of the gas}

The wealth of molecular hydrogen lines detected in both MIR and NIR spectra allows us to perform a detailed study of the H$_2$ excitation in a high-mass jet, taking into account, for the first time, also the pure rotational lines. 

As a result, employing all the available H$_2$ ratios in our ro-vibrational diagrams, 
we have derived column densities, extinction, and temperature of the gas.
Combining these parameters with the 2.12\,$\mu$m flux obtained from the narrow-band imaging, 
we have determined an accurate measurement of the H$_2$ luminosity ($L_{H_2}$) for each knot and for the entire
flow (see Caratti o Garatti et al.~\cite{caratti} for a detailed description of this procedure).

\subsubsection{Ro-vibrational diagrams from NIR lines}
\label{ro-vib:sec}
As a first step of our analysis, we employed only the NIR lines. 
Line pairs originating from the same energy level should lie in the same position
of the ro-vibrational diagram: by varying the extinction value ($A_{\rm v}$) and increasing the 
goodness of the fit (maximising the correlation coefficient) 
extinction and temperature can be evaluated simultaneously 
(see e.\,g. Giannini et al.~\cite{giannini04}, Davis et al.~\cite{davis04}, Caratti o Garatti et al.~\cite{caratti}).
In order to reduce the uncertainties, only the transitions with a S/N$>$3 
and not affected by blending with other lines were used.
The result of this analysis is shown in Table~\ref{phys:tab}, where the average excitation temperature,
the $A_{\rm v}$, and the column densities of the warm gas component are listed for each knot (columns 2--4).
The ro-vibrational diagrams are shown in Fig.~\ref{rotall:fig}.

To compute the extinction in each knot, we have selected all the available pairs of lines.
From the spectrum of knot\,C 41 lines have been used, 
giving 29 different ratios (i.\,e. the ratios are among the following groups of transitions: 1-0S(i),1-0Q(i+2);
2-0S(i),2-0Q(i+2),2-0O(i+4),2-1S(i); 3-1S(i),3-1Q(i+2),3-1O(i+4); with i=0,1,2...).
The best fit was obtained excluding the 1-0S(i)/1-0Q(i+2) pairs,
that deviated more than 3$\sigma$ from the average value. The final value is $A_{\rm v}$=7.6$\pm$0.2,
and has a very small error. For the other knots, the $A_{\rm v}$ ranges from 6 to 10\,mag, 
with errors between 1 and 3\,mag (see Tab.~\ref{phys:tab}, column 3). For these knots,
a few pairs of lines were available and hence the errors are larger.

With the exception of knot C, all the lines in 
our ro-vibrational diagrams are well fitted by a single straight line (see Fig.~\ref{rotall:fig}),
indicating a uniform temperature of the gas. The observed temperatures range from 2000 to 2500\,K.

On the other hand, the different H$_2$ lines detected in knot C cannot be fitted by a single line.
The ro-vibrational diagram exhibits a typical curvature indicating the presence of a stratification in the gas temperature.
A more elaborate model of a mixture of gas at two different temperatures can describe these data (see e.\,g. Giannini et al.~\cite{giannini02}, Caratti o Garatti
et al.~\cite{caratti}, Gredel~\cite{gredel07}).
The population densities of the lines coming from levels with an excitation energy up to $\sim$12\,000\,K 
are consistent with an excitation temperture of 2050\,K. These lines indeed trace the warm component of the gas. 
Conversely, lines with a higher excitation energy ($>$12\,000\,K) are thermalised at a higher temperture of $\sim$5200\,K.
These lines trace the hot component of the gas. A single fit through all the lines gives only a measure of the `averaged' 
temperature ($\sim$3300\,K) (see Fig.~\ref{rotall:fig}, top left panel).
In our model the hot component is a fraction of the gas of about 8\%.

Another important parameter that derives from ro-vibrational diagrams is the column density of the gas.
The column densities of the warm H$_2$ component are reported in Table~\ref{phys:tab}. They are very similar to those measured in other high-mass protostellar jets (10$^{18}$-10$^{19}$\,cm$^{-2}$) (see e.\,g. Davis et al.~\cite{davis04}, Gredel~\cite{gredel}), and they are 1-2 orders of magnitude larger than the values observed in low-mass jets.

Finally, from the analysis of Fig.~\ref{rotall:fig}, it is worth to note that the H$_2$ gas along the flow is fully thermalised. The fact that we detect no lines with v$\ge$6 indicates that fluorescence mechanisms do not play an important role in the excitation (see e.\,g. Black \& van Dishoeck~\cite{black}). Excitation by non-thermal processes like fluorescence in presence of a UV field would cause the vibrational temperature to be different from the rotational temperatures, and thus lines belonging to different vibrational series should not be located on the same line (see e.\,g. Hora \& Latter~\cite{hora}). As a consequence, we should observe strong deviations from the smooth (or linear) Boltzmann distribution of our ro-vibrational diagrams (see also Gredel~\cite{gredel07}). 
Apparently, this contrasts with results of Sec.~\ref{lowresISO:sec}, where the presence of a PDR was inferred.
Such a PDR could produce enough far-UV radiation to excite the higher vibrational levels of the H$_2$ 
and induce a fluorescent emission (see e.\,g. Burton~\cite{bur}), that we do not detect.
To evaluate a possible contribution of the PDR to the observed lines,
we used the `PDR toolbox'\footnote{The `PDR Toolbox' is available at http://dustem.astro.umd.edu and contains downloadable FIR line diagnostic information about PDRs. The tool has been created by L. Mundy, M. Wolfire, S. Lord, and M. Pound, and it is based on the new PDR models of Kaufman et al.~(\cite{kauf}).}. 
The observed [\ion{C}{ii}] 158\,$\mu$m diffuse line intensity (considering a LWS beam-width of 80$\arcsec$) is
$\sim$2.5$\times$10$^{-4}$\,erg\,s$^{-1}$\,cm$^{-2}$\,sr$^{-1}$. Assuming that all the emission arises from the PDR and
that the density is 10$^5$-10$^6$\,cm$^{-2}$ (Cesaroni et al.~\cite{cesaroni99}), we obtain a relatively faint FUV flux 
of $G_0$ between 10 and 100 (where $G_0$ is measured in units of 1.6$\times$10$^{-3}$\,erg\,s$^{-1}$\,cm$^{-2}$).
Under these circumstances the PDR surface temperture does not exceed 100\,K (Kaufman et al.~\cite{kauf}) and the contribution
to the overall H$_2$ emission is negligible. For example, the diffuse line intensity of the 0-0\,S(1) is between 10$^{-6}$ and 10$^{-5}$\,erg\,s$^{-1}$\,cm$^{-2}$\,sr$^{-1}$, while from our observations we obtain 
$\sim$6$\times$10$^{-4}$\,erg\,s$^{-1}$\,cm$^{-2}$\,sr$^{-1}$ (dividing by the ISO-SWS FoV). In the NIR
the emission from the shock (at 2.12\,$\mu$m) is four orders of magnitude larger. These results
clearly indicate that the detected H$_2$ mostly arises from shocks.

\subsubsection{The cold H$_2$ component from MIR lines}
As a second step in our analysis, we also used the ISO-SWS lines to study the cold component of the gas,
that is usually traced by lines with excitation energy lower than 5000\,K, namely the 0-0\,S lines (with J$_{fin}\le$5) 
between 6 and 28\,$\mu$m in the ISO-SWS spectrum. 

\begin{figure*}
 \centering
 \includegraphics [width=14 cm]{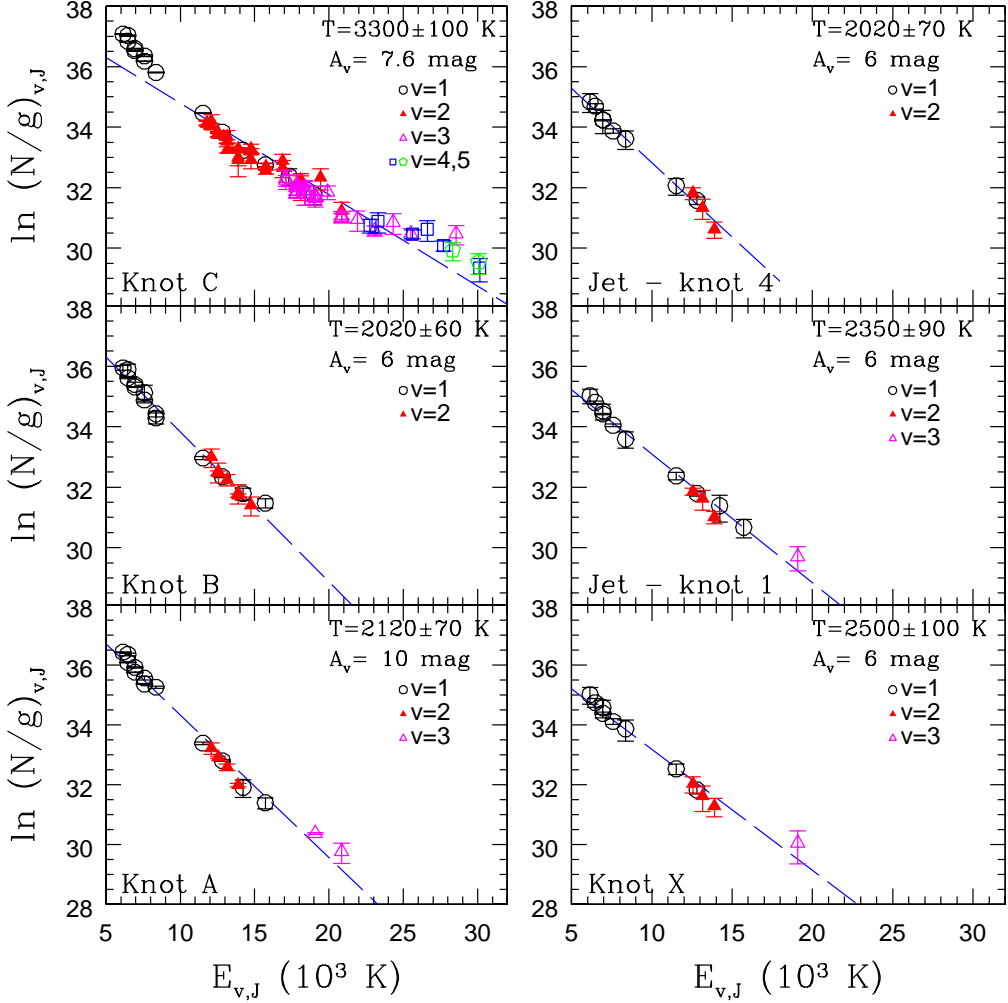}
  \caption{ Rotational diagram of knots in the IRAS\,20126+4104 jet obtained from the low resolution NIR spectroscopy.
   Different symbols indicate lines coming from different vibrational levels, as coded in the upper right corner of
   the boxes. The inferred temperature and extinction is indicated in the upper right corner of the box. 
\label{rotall:fig}}
\end{figure*}

\begin{figure}
 \centering
 \includegraphics [width=9.2 cm]{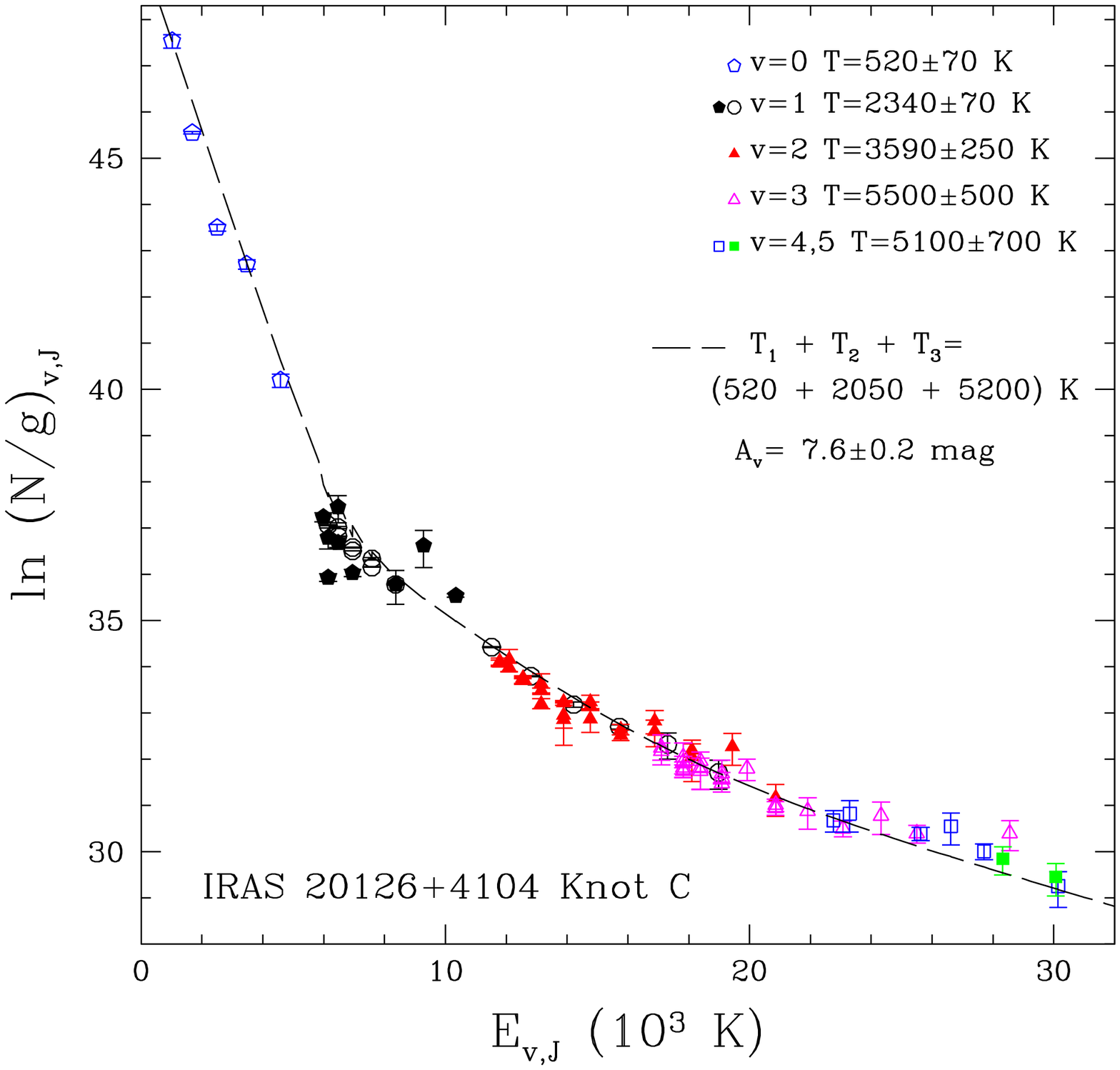}
  \caption{ Rotational diagram of knot C of the IRAS\,20126+4104 jet obtained from the low resolution spectroscopy.
   Different symbols indicate lines coming from different vibrational levels, as coded in the upper right corner of
   the box (v=0 and v=1 ISO lines are indicated by open and filled pentagons, respectively). The dashed line represents the theoretical population distribution for a combination of thermalised gas at three different temperatures (520, 2050, and 5200\,K, see text). The derived temperature from the different vibrational levels is indicated in the upper right corner of the box. The extinction is reported as well. 
\label{rotknotC:fig}}
\end{figure}

Since we do not know the spatial extent of the emitting region, we have matched the column densities of 
the 1-0\,Q lines present in both NICS and SWS spectra, to intercalibrate the data in the ro-vibrational diagram.
Indeed the measured ISO-SWS fluxes of the 1-0\,Q lines are very close to those measured in knot\,C. 
Therefore we have assumed that its column density is representative of the warm H$_2$ component detected with ISO, as well.
The result is shown in Fig.~\ref{rotknotC:fig}. SWS lines coming from the v=0 and v=1 levels are represented as open and filled
pentagons, respectively. The data are overplotted on the ro-vibrational diagram of knot C (shown in Fig.~\ref{rotall:fig}, top left panel). Although there is some scatter among the v=1 data points (two 1-0\,Q and 1-0\,O lines have slightly larger and smaller column densities, respectively), the overall fit is quite satisfactory. In particular, the 0-0\,S lines reveal 
the presence of a cold gas component at about 520\,K, that has a larger column density than the other components ($N_{H_2}\sim$9.7$\times$10$^{21}$\,cm$^{-2}$, see also Fig.~\ref{rotknotC:fig}). 
 
In order to reproduce the observed ro-vibrational diagram, we calculated a theoretical H$_2$ spectrum for a mixture
of three H$_2$ layers in LTE condition at a temperature of 520, 2050, and 5200\,K. The adopted LTE code (see also Caratti o Garatti et al.~\cite{caratti}) computes the line intensities involving levels with $0\le$v$\le14$ and $0\le J \le29$ (E$_{v,J}\le$ 50\,000\,K). Ro-vibrational energies are taken from Dabrowsy~(\cite{dab}) and the Einstein coefficients from Wolniewicz et al.~(\cite{wol}). We have assumed an ortho/para ratio equal to three (as also evinced from our ro-vibrational diagrams).
The resulting model is plotted as a dashed line on the points of Fig.~\ref{rotknotC:fig}, and properly fits our observations. 
In this case the warm and hot components are only a small fraction of the total H$_2$ gas, i.\,e. less than 1\%.

\subsubsection{H$_2$ luminosity}
In the last step of our analysis we inferred $L_{H_2}$.
Once the physical parameters ($A_{\rm v}$, $T$) for each knot were derived, we dereddened the 2.12\,$\mu$m flux obtained 
from the imaging (see Tab.~\ref{phys:tab}, column 5), adopting the Rieke \& Lebofsky~(\cite{riek}) reddending law, 
and computed the line ratios with the other H$_2$ lines by applying our radiative LTE code at $T=T_{avg}$. 
From these ratios, the absolute intensities of individual lines were computed and the H$_2$ luminosity of each knot
was derived (see Tab.~\ref{phys:tab}, column 6). The sum of these values gives a total $L_{H_2}$ of the flow of 4.6$\pm$0.3\,$L_{\sun}$. 
An average temperature reproduces the contribution of both the warm and hot gas components to the radiated energy
in H$_2$ well (see Caratti o Garatti et al.~\cite{caratti}).
We are not taking into account the luminosity of the `cold' component, that, however, due to the extremely high 
column density of the gas, here can be considerable. 
In order to evaluate this, we have assumed that the measured SWS flux from the 0-0\,S lines is the total flux emitted by the shocked cold H$_2$ gas of the flow, and that the contribution of non-thermal emission is negligible, as seen in Sec~\ref{ro-vib:sec}.
We then computed the luminosity of an LTE gas at $T$=520\,K, using the 0-0\,S(1) observed dereddened intensity to derive the absolute intensities of the remaining lines. The luminosity of the `cold' gas is 3.6$\pm$0.6\,$L_{\sun}$. Added to the previous estimate this gives a total $L_{H_2}$ for the entire flow of 8.2$\pm$0.7\,$L_{\sun}$.

\begin{table*}
\caption[]{Physical parameters of the H$_2$ knots in IRAS\,20126+4104 derived from the low resolution spectroscopy and imaging.
\label{phys:tab}}
\begin{center}
\begin{tabular}{ccccccc}
\hline \hline\\[-5pt]
Knot &  Average Temperature & $A_{\rm v}(H_2)$ & $N_{H_2}$ & Flux (2.12\,$\mu$m)  &  $L_{H_2}^a$  &   $\dot{M}_{out}$(H$_2$)$^b$\\
     &    (K)       & (mag) & 10$^{18}$cm$^{-2}$ & (10$^{-14}$erg\,cm$^{-2}$\,s$^{-1}$) & ($L_{\sun}$) & (10$^{-8}$ $M_{\sun}$\,yr$^{-1}$) \\
\hline\\[-5pt]
A                    &    2120$\pm$70  &  10$\pm$3     & 5.4       & 29.9$\pm$0.2       & 0.70$\pm$0.13   & 18          \\
B                    &    2020$\pm$60  &  6$\pm$1$^c$  & 3.6       & 10.9$\pm$0.1       & 0.21$\pm$0.03   & 6          \\
C                    &    3300$\pm$100 & 7.6$\pm$0.2   & 12.5      & 71.5$\pm$0.3       & 3.12$\pm$0.24   & 192          \\
D                    &    $\cdots$     &   $\cdots$    & $\cdots$  & 6.4$\pm$0.1        & 0.12$^d$        & $\cdots$     \\
X                    &    2500$\pm$100 &   6$\pm$1$^e$ & 1.9       & 2.4$\pm$0.1$^f$    & 0.06$\pm$0.01   & 2            \\
jet - knot 1         &   2350$\pm$90   &   6$\pm$3     & 1.7       & 7.5$\pm$0.2        & 0.2$\pm$0.1     & 8            \\
jet - knot 2         &   $\cdots$     &   $\cdots$    & $\cdots$   & 1.0$\pm$0.4        & 0.02$^g$        & $\cdots$     \\
jet - knot 3         &   $\cdots$     &   $\cdots$    & $\cdots$   & 2.7$\pm$0.5        & 0.04$^h$        & $\cdots$     \\
jet - knot 4         &     2020$\pm$70 &   6$\pm$3     & 1.2       & 10.0$\pm$0.2       & 0.2$\pm$0.1     & 8            \\
\hline \hline
\end{tabular}
\end{center}
Notes: 
$^a$ Derived from the average temperature. The contribution from the cold component of the gas is not taken into account.
The $L_{H_2}$ from the cold component, as measured from the 0-0\,S lines of the ISO spectrum, is $L_{H_2}$=3.6$\pm$0.4 $L_{\sun}$.\\ 
$^b$ Derived from the average temperature. The total $N_{\rm H2}$ column density is measured by comparing the intrinsic (per unit mass) and observed flux of the 1-0\,S(1) (2.12 $\mu$m) line.\\
$^c$ $A_{\rm v}$ measured from the [\ion{Fe}{ii}] lines (1.64 and 1.26\,$\mu$m) is 11$\pm$7 mag.  \\
$^d$ Computed assuming $T_{ex}$=2000\,K and $A_{\rm v}$=6 mag. \\
$^e$ $A_{\rm v}$ measured from the [\ion{Fe}{ii}] lines (1.64 and 1.26\,$\mu$m) is 7$\pm$4 mag.  \\
$^f$ Flux([\ion{Fe}{ii}]) = 3$\pm$1$\times$10$^{-15}$erg\,cm$^{-2}$\,s$^{-1}$ \\
$^g$ Computed assuming $T_{ex}$=2350\,K and $A_{\rm v}$=6 mag. \\
$^h$ Computed assuming $T_{ex}$=2020\,K and $A_{\rm v}$=6 mag. \\
\end{table*}

\subsection{Mass flux, mass, and energy of the H$_2$ jet}
\label{massflux:sec}
The physical and kinematical parameters inferred in the previous sections allow us to
evaluate important kinematical and dynamical properties of the flow.

Therefore, we have estimated the mass flux rate of the flow ($\dot{M}_{out}$) from the H$_2$ in order to compare it with the mass outflow rate, and the mass accretion rate of the protostar previously obtained from CO observations and models 
(see e.\,g. Shepherd et al.~\cite{shepherd}, Cesaroni et al.~\cite{cesaroni05}, Lebr\'{o}n et al.~\cite{lebron}).

$\dot{M}_{out}$(H$_2$) can be written as $\dot{M}_k = 2 \mu m_H N_{H_2} A  v_t / l_t$, where $\mu$ is the average atomic weight, $m_{H}$ the proton mass, $N_{H_2}$ the H$_2$ column density, A the area of the H$_2$ knot, $v_{t}$ the tangential velocity,
and $l_{t}$ the projected length of the knot (see, e.\,g., Nisini et al.~\cite{nisini05}, Podio et al.~\cite{podio}, 
Antoniucci et al.~\cite{antoniucci}).

Using the 2.12\,$\mu$m column density of Tab.~\ref{phys:tab}, the radial velocities of Tab.~\ref{H2vrad:tab}, 
and assuming an average inclination for all the knots of 20$\degr$ with respect to the plane of the sky, we estimate mass fluxes
between 10$^{-6}$ and 10$^{-8}$\,$M_{\sun}$\,yr$^{-1}$ (see Tab.~\ref{phys:tab}, column 7). In fact, these values represent lower limits. 
From the ro-vibrational analysis the cold gas column density is two orders of magnitude larger than the warm component
(e.\,g. 0-0\,S lines have a column density $N_{H_2}\sim$9.7$\times$10$^{21}$\,cm$^{-2}$).
Using this value, and assuming the same velocity and extension of the emitting area of the 2.12\,$\mu$m line,
we obtain $\dot{M}_{out}$(H$_2$)$\sim$7.5$\times$10$^{-4}$\,$M_{\sun}$\,yr$^{-1}$.

It is worth noting, however, that this value should only be considered as an estimate of the mass flux, for two reasons. 
Firstly, we have set $l_{t}$ and $v_{t}$ equal for both H$_2$ components.
Probably, such an estimate is a lower limit, since $l_{t}$ of the cold gas can be larger than the $l_{t}$ measured at 2.12\,$\mu$m. 
More important, the adopted mass flux formula lays on the assumption that $l_{t}/v_{t}$ is the \textit{cooling time} ($t_c$) of the shock.
In this case, the kinematically determined $t_c$ of knot C is 1.9$\times$10$^{10}$\,s.
Also other methods have been used in the literature to derive $t_{c}$: Davis et al.~(\cite{davis00}) adopt the approximation 
$t_{c} \sim 3 \times 10^{8}n_{6}^{-1}T_{3}^{-2.3}$ (in seconds), where $n_6$ is the density in units 10$^6$\,cm$^{-3}$ and $T_3$ is the temperature in units of 1000\,K (Smith \& Brand~\cite{sb}). In this way, adopting $n_{H_2}$=10$^5$\,cm$^{-3}$ (Cesaroni et al.~\cite{cesaroni99}) and two different temperatures, for knot C, we would obtain 
$t_{c}$(warm)= 6.1$\times$10$^{8}$\,s (with $T$=2000\,K), and $t_{c}$(cold)= 1.5$\times$10$^{10}$\,s (with $T$=500\,K).
The $\dot{M}_{out}$ of the cold component would then be remarkably similar to the previous estimate. Moreover, it is
very close to the mass outflow rates derived by Shepherd et al.~(\cite{shepherd}) and Lebr\'{o}n et al.~(\cite{lebron}) from the CO mm analysis (8.1$\times$10$^{-4}$\,$M_{\sun}$\,yr$^{-1}$, and 3.4$\times$10$^{-3}$\,$M_{\sun}$\,yr$^{-1}$, respectively).

In addition, we can derive a further estimate of the mass flux from the luminosity of the [\ion{O}{i}] line (at 63\,$\mu$m)
using the approximate formula $\dot{M}_{\rm out}\sim$10$^{-4}\times$(L(63\,$\mu$m)/$L_{\odot}$)\,$M_{\sun}$\,yr$^{-1}$ (see e.\,g. Hollenbach \& McKee~\cite{hollen}, Liseau et al.~\cite{liseau}, Cabrit~\cite{cabrit}). As a result, we get $\dot{M}_{\rm out}$(OI) = 2$\times$10$^{-4}$\,$M_{\sun}$\,yr$^{-1}$. Such an estimate is, however, an upper limit, since part or all of the [\ion{O}{i}] emission
could originate from the PDR.

To derive the mass of each knot, we assume $M_k = 2 \mu m_H m_H N_{H_2} A$.
For knot D, where $N_{H_2}$ was not measured, we have assumed an average value of 4.4$\times$10$^{18}$\,cm$^{-2}$ from Tab.~\ref{H2vrad:tab}. 
The total mass ($\Sigma_{k}M_k$) of the warm gas is $\sim$10$^{-3}$\,$M_{\sun}$. 
Assuming an average value for the inclination of the flow of 20$\degr$, the dynamical timescale of the flow ($\tau_d$) is $\sim$1.3$\times$10$^{4}$\,yr, the total momentum $P$($\Sigma_{k}M_k$v$_k$) is 0.06\,$M_{\sun}$\,km\,s$^{-1}$, the kinetic energy $E_k$($\frac{1}{2}\Sigma_{k}M_k$v$_k^2$)
is 5$\times$10$^{43}$\,ergs, and the momentum flux $\dot{P}$ ($P$/$\tau_d$) is 5$\times$10$^{-6}$\,$M_{\sun}$\,yr$^{-1}$\,km\,s$^{-1}$.
We note that these values must be considered as lower limits, since the mass of observable shocked warm H$_2$ is
only a small fraction of the total shocked molecular hydrogen mass, as seen previously.
If we compute these quantities taking into account the column density of the cold component of the gas, we obtain the 
following crude estimates: $M\sim$0.6\,$M_{\sun}$\,km\,s$^{-1}$, $P\sim$50\,$M_{\sun}$\,km\,s$^{-1}$, $E_k\sim$4$\times$10$^{46}$\,ergs, 
$\dot{P}\sim$4$\times$10$^{-3}$\,$M_{\sun}$\,yr$^{-1}$\,km\,s$^{-1}$.

Finally, we can give a very rough estimate of the density for the cold component, as well. If we assumed that
the size for knot C along the line of sight was similar to the observed length ($\sim$10$^{17}$\,cm), we would obtain 
$n_{H_2}\sim$10$^5$\,cm$^{-3}$, as in Cesaroni et al.~(\cite{cesaroni99}).

\section{Discussion}
\label{discussion:sec}
\subsection{A precessing jet model}
\label{prec:sec}

\begin{figure}
 \centering
   \fbox{\includegraphics [width=8.0 cm] {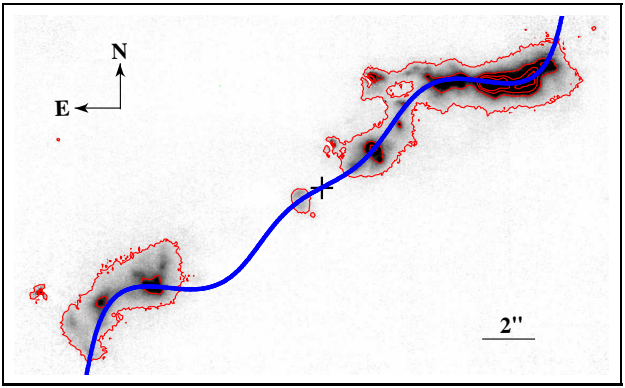}}
   \caption{ Precession model plotted over the H$_2$-Br$\gamma$ image from Subaru. The model is for a precession angle of 7\fdg6 and
a precession scale of 10\farcs9. A cross indicates the position of the source.
\label{prec:fig}}
\end{figure}

As mentioned in Sect.~\ref{ima_res:sec}, the inner region of the H$_2$ jet clearly shows a wiggling morphology, possibly indicating
a small-scale precession, different from the long-period precession derived from the positions of the outer knots 
(Shepherd et al.~\cite{shepherd}, Cesaroni et al.~\cite{cesaroni05}).

In order to investigate this new morphological feature, we used a simple jet model to fit the observed position of 
the inner knots (see e.\,g. Eisl\"{o}ffel et al.~\cite{eis}, Texeira et al.~\cite{tex}):
\begin{equation}
\label{model:eq}
\left(\begin{array}{cc}     \alpha \\      \delta \end{array} \right) =
\left(\begin{array}{cc}     \alpha_0 \\         \delta_0 \end{array} \right) +
\left(\begin{array}{cc}   cos(\psi) & -sin(\psi) \\     sin(\psi) & cos(\psi) \end{array} \right) \times
\left(\begin{array}{cc}    \phi l sin(2 \pi l / \lambda + \chi_0 ) \\      l \end{array} \right)
\end{equation}
where $\alpha_0$, $\delta_0$ are the coordinates of the source, $\psi$ the rotation angle, $\phi$ the precession amplitude (in radiants), $l$ the distance from the source position, $\lambda$ the precession length scale, $\chi_0$ the initial phase at the source.
The resulting model, shown in Fig.~\ref{prec:fig}, has been obtained for values of $\psi$= -60\fdg9, $\phi$=0.133 ($\sim$7\fdg6), 
$\lambda\sim$10\farcs9, $\chi_0$=81\fdg6. The inclination of the flow with respect to the plane of the sky was not considered in the fit, since its value (close to the source) is relatively small ($\sim$9$\degr$). 
For a constant jet velocity of 80\,km\,s$^{-1}$, at a distance ($D$) of 1.7\,kpc, we obtain a precession period of $\sim$1100\,yr, that appears short in comparison with the value of $\sim$64\,000\,yr derived in Shepherd et al.~(\cite{shepherd}) and Cesaroni et al.~(\cite{cesaroni05}). 
We have then tried to reproduce the complex motion of the flow, adding a second precession mode 
to our model, that should take into account the longer period traced by the outer knots. 
We were, however, not able to fairly match the pattern of the entire jet. 

This could indicate that the origin of the two observed `precessions' is different,
or that the precession period changed over time. 
The long-period precession was interpreted by Shepherd et al.~(\cite{shepherd}) and Cesaroni et al.~(\cite{cesaroni05}) as caused by the interaction between the disc of IRAS\,20126+4104 and a stellar companion of a few solar masses.
However, it is unlikely that the small-period precession we observe is due to the orbital motion of the jet source around its companion.
Since the orbital radius is given by the following equation (see Anglada et al.~\cite{anglada}):
\begin{equation}
r=\frac{\lambda tan(\phi) D}{2\pi}
\end{equation}
the resulting value is $r\sim$400\,AU. The radius of IRAS\,20126+4104 disc was estimated around 800\,AU 
on the contrary (Cesaroni et al.~\cite{cesaroni97}), meaning that the companion would intersect the disc.
Alternatively, the precession could have changed with time. The ejection of a third companion in a hierarchical triple 
system, for example, might have led to the formation of a tighter binary system and to a shortening of the orbital period. However, this hypothesis does not explain the different position angles on the sky measured in the two precession modes (-60\fdg9 and -37$\degr$, respectively).

It seems more likely, then, that the complex axis wandering is induced by tidal interactions of multiple stellar companions. 
Indeed IRAS\,20126+4104 is not a single compact source, but it is composed by a small cluster of YSOs (De Buizer~\cite{debuizer}), therefore the dynamics of such a system can be extremely complicated.
It is indeed beyond of the scope of this paper to obtain a rigorous modelling for the dynamic of such a system.

\subsection{H$_2$ jet vs CO outflow}
\label{H2vsCo:sec}

Observations of protostellar jets in low-mass YSOs support the idea
that their molecular outflows are driven and accelerated by the highly-collimated bipolar jets. 
Outflows from intermediate- and high-mass sources, on the other hand, are often poorly collimated
and only in a few cases there is clear evidence of a driving jet. This has often led
to the suggestion that these outflows could be driven by a different mechanism, as, e.\,g., a wide-angle 
radial wind (see e.\,g. Arce et al.~\cite{arce}).

Apparently, IRAS\,20126+4104 represents a particular example of poorly collimated outflow powered by a protostellar jet.
The poor collimation of this outflow has been explained by means of the severe precession of the jet.
It is thus interesting to investigate if the properties of the IRAS\,20126+4104 H$_2$ jet meet the physical parameters of the
CO outflow, i.\,e. if the jet is powerful enough to accelerate the surrounding medium and drive the outflow.
This is indeed a fundamental question, because the YSO accretion rate is often estimated from the mass flux rate of the outflow.
Comparing CO and H$_2$ parameters, however, we have to keep in mind that the CO
lines only give a time-integrated response over the lifetime of the jet, whereas the H$_2$ emission provides an `instantaneous' 
measure of these quantities.

Large-scale outflow properties from CO mm observations were extensively studied by several authors 
(see e.\,g. Shepherd et al.~\cite{shepherd}, Lebr\'{o}n et al.~\cite{lebron}). Since their computations
are based on different assumptions, however the results of Lebr\'{o}n et al.~(\cite{lebron}) are at least one 
order of magnitude larger than those of Shepherd et al.~(\cite{shepherd}).

\begin{table}
\caption[]{ Comparison between H$_2$ jet and CO outflow physical properties of the IRAS\,20126+4104 flow.
\label{physcomp:tab}}
\begin{center}
\begin{tabular}{cccc}
\hline \hline\\[-5pt]
Parameter &  H$_2$ &    CO$^a$ & CO$^b$ \\
\hline\\[-5pt]
Mass ($M_{\sun}$)   &   0.6      &    53  &  16.3             \\
v$_{red}$ (km\,s$^{-1}$) &   (-8, 47)      &  $\cdots$  &  (4, 55)\\
v$_{blue}$ (km\,s$^{-1}$) &   (-14, -42)      &  $\cdots$  &  (-11, -58)\\
$\tau_d$ (10$^{4}$\,yr) & 1.3 & 6.4 & $\cdots$ \\
$\dot{M}$ (10$^{-3}$\,$M_{\sun}$\,yr)   & 0.75 & 0.81 & 3.4 \\
E$_k$(10$^{46}$\,ergs) & 4 & 5.1 & 130 \\
P ($M_{\sun}$\,km\,s$^{-1}$) & 50 & 403 & 1490 \\
$\dot{P}$ (10$^{-3}$\,$M_{\sun}$\,yr$^{-1}$\,km\,s$^{-1}$) & 4 & 6 & 310 \\
\hline \hline
\end{tabular}
\end{center}
Notes: \\
$^a$ From Shepherd et al.~(\cite{shepherd}).\\ 
$^b$ From Lebr\'{o}n et al.~(\cite{lebron}).\\
\end{table}

In Table~\ref{physcomp:tab} we compare our H$_2$ flow parameters (obtained including the cold component) with 
the parameters of the CO outflow from Shepherd et al.~(\cite{shepherd}) and Lebr\'{o}n et al.~(\cite{lebron}). 
From this table, we can infer that the jet is, at least partially, driving the outflow.
The radial velocities of H$_2$ are well consistent with the CO velocities, including the high velocity component up to 60\,km\,s$^{-1}$
detected by Lebr\'{o}n et al.~(\cite{lebron}). On the other hand, the total mass of the H$_2$ bullets is just a fraction of 
the total CO mass. As a consequence, the momentum ($P$) of the jet is at least one order of magnitude smaller than that of the outflow. 
However, our estimates of the mass flux, kinetic energy, and momentum flux are, indeed, almost coincident with Shepherd et al.~(\cite{shepherd}),
supporting the hypothesis that the H$_2$ jet is fully driving the flow. Contrarily, the values from Lebr\'{o}n et al.~(\cite{lebron})
are from one to two orders of magnitude larger than ours (see Tab.~\ref{physcomp:tab}). This would imply that the 
H$_2$ jet is only partially driving the outflow, and that there is an undetected component along the flow, likely a neutral jet, that is
 supplying further momentum flux to the outflow. However, it is worth to note that the mass flux obtained by Lebr\'{o}n et 
al.~(\cite{lebron}) is slightly larger than the mass accretion rate of IRAS\,20126+4104 inferred by Cesaroni et al.~(\cite{cesaroni05}) 
(2$\times$10$^{-3}$\,$M_{\sun}$\,yr$^{-1}$), and this is quite unlikely. This discrepancy could be then explained by the presence of 
(undetected) multiple jets (winds), driven by other YSOs near the source, that would substantially contribute to power the CO outflow. 
 
\subsection{Shock conditions along the jet}
\label{conditions:sec}
As the previous results point out, the cold H$_2$ component of the jet plays a major role in the kinematics and dynamics 
of this outflow. It is therefore desirable to discuss the origin of this component in more detail.

A prevalence of C-type shocks along the flow could explain our findings.
Indeed C-type shocks produce a high column density in the 0-0 lines (see e.\,g. Flower et al.~\cite{flower}, McCoey et al.~\cite{mccoey}), that are very sensitive to these shocks. The column densities in our 0-0 lines could not be explained by a J-type shock (see e.\,g. Le Bourlot et al~\cite{lebour}).
Moreover, except for the [\ion{Fe}{ii}] and the [\ion{O}{i}] emission close to the source, the emission observed along the flow comes only from the H$_2$. Only the inner region near the source position exhibits hints of ionic emission. There is no evidence of the
ionised jet detected in the radio by Hofner et al.~(\cite{hofner}), probably confined to a region very close to the source and highly extincted. On the other hand, the complete absence of any ionic emission in the other knots of the flow should not be attributable to a high value of extinction, since the measured $A_{\rm v}$ values along the jet are relatively small. Such a behaviour has already been observed in several low-mass protostellar jets.
They usually show only H$_2$ emission, or, at most, a hint of ionic emission close the exciting source (see e.\,g. Giannini et al.~\cite{giannini04},
Caratti o Garatti et al.~\cite{caratti}). It is also well observed in the high-mass IRAS\,18151-1208 jet spectra (Davis et 
al.~\cite{davis04}). They are interpreted as C-type shocks, or as J-shocks with magnetic precursors, that is J-type shock waves evolving into "continuous'' C-type shock waves (see e.\,g. Giannini et al.~\cite{giannini04}). 
Also our kinematical observations seem to indicate the presence of C-type shocks. H$_2$ velocities in C-type shocks can exceed 50-60\,km\,s$^{-1}$ up to 80\,km\,s$^{-1}$ (see e.\,g. Le Bourlot et al.~\cite{lebour}).
Accordingly, the \textit{FWZI} of H$_2$ bow-shocks can be up to twice this value, depending on the inclination of the jet axis with respect to the plane of the sky (see e.\,g. Davis et al.~\cite{davis04}), 
that is the broadest lines are expected for bow-shock located in the plane of the sky.
Actually, if we consider the different inclinations for the inner and the outer knots, our observations seem to indicate a high shock velocity in both regions, whereas the larger value of the \textit{FWZI} is observed in the inner knots (closer to the sky plane), and a less broad shape in the outer knots. From these values we can roughly estimate a shock velocity between 40 and 
80\,km\,s$^{-1}$, where the fastest speed would be attained in the inner region.
Such high shock velocities (50-80\,km\,s$^{-1}$) however imply low pre-shock densities of 10$^3$-10$^4$\,cm$^{-3}$ (Le Bourlot et al.~\cite{lebour}) in C shocks, assuming a magnetic field $B$($\mu$G)=b[$n_H$(cm$^{-3}$)]$^{-0.5}$, with b=1. 
Higher densities lead to a lower maximum shock speed (v$_{diss}$), that is the maximum shock velocity which can be attained, prior to the collisional dissociation of H$_2$. On the contrary, both CO and H$_2$ observations seem to indicate a higher pre-shock density (10$^5$-10$^6$\,cm$^{-3}$) in the medium
(see e.\,g. Cesaroni et al.~\cite{cesaroni99}). In this case v$_{diss}$ would decrease to 30-50\,km\,s$^{-1}$ (Le Bourlot et al.~\cite{lebour}). Thus a `standard' C-type model cannot entirely explain our findings close to the source. A larger transversal magnetic field (b$>$1) or a different shock model (as a J-shock with a magnetic precursor) are probably needed.
This last hypothesis would also explain the presence of the ionic emission in the knots close to the source.

\subsection{$L_{H_2}$, accretion and ejection rates}
\label{LH2:sec}
\begin{figure}
 \centering
   \includegraphics [width=9.2 cm] {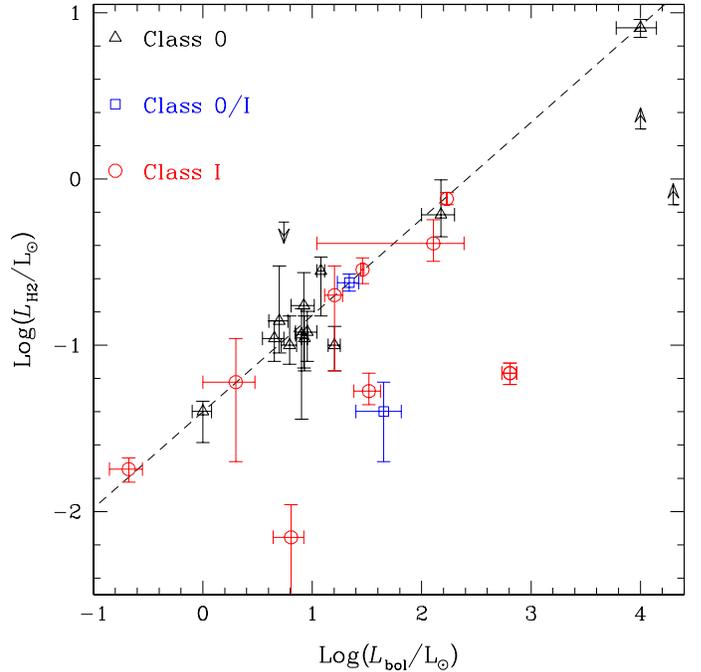}
   \caption{ $L_{H_2}$ vs $L_{bol}$ including the $IRAS 20126+4104$ jet (also considering the luminosity of the cold component, the new datapoint is located in the upper-right corner of the diagram). Values for IRAS\,18151-1208 and IRAS\,11101-5829 jets have been included
as lower limits, as well (see discussion in the text). A dashed line indicates the previous fit from Caratti o Garatti et al.~(\cite{caratti}).
\label{LH2:fig}}
\end{figure}

Several authors have argued that a tight correlation between the evolutionary properties of YSOs and their outflows would exist.
This is particularly true for those objects forming by disc accretion, since accretion and ejection should be regulated by the same mechanism.
Recently, it has been shown for low-mass YSOs that the total H$_2$ luminosity of the jets is proportional to their accretion 
rates (see e.\,g., Smith~\cite{smith}, Froebrich et al.~\cite{froebrich03}). In particular, an empirical relationship between $L_{H_2}$ 
and $L_{bol}$ ($L_{H_2}$ $\propto$ $L_{bol}^{0.58}$, Caratti o Garatti et al.~\cite{caratti}) was derived for a large sample of 
low-mass protostellar jets.
This correlation holds for very young YSOs (Class\,0, and some Class\,I), where the bolometric luminosity is mostly coincident with the 
accretion luminosity of the object, and implies that $\dot{M}_{acc}$ increases with the luminosity (i.\,e. mass) of the protostar.
In Fig.~\ref{LH2:fig} we report the results obtained from the sample of 
Caratti o Garatti et al.~(\cite{caratti}). The diagram compares the measured outflow H$_2$
luminosity versus the bolometric source luminosity, both on a logarithmic scale.
We have included the new data from IRAS\,20126+4104 (positioned in the upper right corner).
The dashed line indicates the best fit previously obtained, that fits perfectly to our object,
suggesting that the relationship applies also to more massive jets in their earliest stage of formation. 
Accordingly to the dynamical timescales of the outflow and the jet (few times 10$^4$ years), 
IRAS\,20126+4104 has not yet reached the main sequence (MS), and has not yet developed any hypercompact HII (HCHII) region,
that may affect the collimation of the jet/outflow system (Beuther \& Shepherd~\cite{beu}).
Most importantly, the bolometric luminosity of the source mainly comes from accretion (see Cesaroni et al.~\cite{cesaroni99}, Cesaroni et al.~\cite{cesaroni05}). In these works, the authors obtain for this source
an accretion luminosity of 1.2$\times$10$^{4}$\,$L_{\sun}$ for a mass accretion rate of $\sim$1-2$\times$10$^{-3}$\,$M_{\sun}$\,yr$^{-1}$.
Moreover, the fact that we obtain an ejection rate of 30-40\% of the accretion rate from the H$_2$ strengthens the reliability of our findings, since for high-mass sources we would expect a ratio of the mass ejection/accretion rate larger than the value ($\sim$0.1) usually derived for the low-mass objects (see e.\,g. Cabrit~\cite{cab07}).

Finally, we compare our results with the $L_{H_2}$ of two high-mass jets previously investigated by means of NIR spectroscopy,
i.\,e. IRAS\,18151-1208 ($L_{H_2}$=0.7\,$L_{\sun}$, Davis et al.~\cite{davis04}) and IRAS\,11101-5829 
($L_{H_2}\ge$2\,$L_{\sun}$, Gredel~\cite{gredel}). These sources have almost the same $L_{bol}$ as IRAS\,20126+4104.
The $L_{H_2}$ estimates are quite close to the H$_2$ luminosity of IRAS\,20126+4104 obtained from the warm H$_2$ component (i.\,e. from our NIR analysis, $L_{H_2}$=4.6$\pm$0.3\,$L_{\sun}$). For comparison, the values of the IRAS\,18151-1208 and IRAS\,11101-5829 jets have also been included as lower limits in Fig.~\ref{LH2:fig}. It is also worth to note that the slightly lower value found by Davis et al.~(\cite{davis04}) could be caused by the high extinction values (10-30 mag) observed towards this flow. 
Indeed a MIR investigation could reveal if the cold H$_2$ component plays a major role in the cooling of those jets, as well.

\subsection{Comparing high- and low-mass jets}
\label{compatison:sec}

The three high-mass protostellar jets spectroscopically investigated up to date share similar characteristics.
As the low-mass jets, they are collimated and powerful enough to drive their outflows. The jet is formed close to the
source, as we observe e.\,g. in IRAS\,20126+4104 knot X (located $\sim$1700\,AU from the source),
or in IRAS\,11101-5829 knot HH136\,J2 (located $\sim$3000\,AU from the source), indicating that the collimation of the
jet occurs close to the source.
They show molecular and ionic shocked emission along the flow, and no evidence of fluorescent 
excitation is detected. The observed jet velocities are similar to those of the CO outflow. 
Where the extinction is low, some jets (e\,g. IRAS\,11101-5829, IRAS\,18162-2048) show HH objects, as well. 
Some kinematical and dynamical quantities as mass flux, momentum flux, luminosity, and kinetic energy, are
larger however, than in low-mass jets, because the powering YSO is more massive.
Moreover, a precession-wiggling morphology is observed in most of the massive collimated H$_2$ jets,
often with large precessing angles (up to $\sim$40$\degr$ in IRAS\,20126+4104)
(e.\,g. IRAS\,16547-4247, Brooks et al.~\cite{brooks}, IRAS\,18151-1208, Davis et al.~\cite{davis04}), IRAS\,11101-5829, Gredel~\cite{gredel}, M17 disc silhouette, N\"{u}rnberger et al.~\cite{nur}, IRAS\,07427-2400, IRAS\,20293+3952, and IRAS\,23033+5951, Nanda Kumar et al.~\cite{nanda}, and IRAS\,20126+4104). 
Precession is also observed in low-mass jets, but it is not as frequent and as pronounced (usually the 
precession angles are less than 10$\degr$).
All this indicates that the dynamical interactions among massive stars are stronger and more frequent than 
in low-mass star forming regions. Such complex dynamics could also explain the confused H$_2$ morphology of some jets
and the lack of collimation in some massive outflows.
Finally, the lack of jets and the poor outflow collimation observed in several massive young sources fits well the evolutionary outflow scenario proposed by Beuther \& Shepherd~(\cite{beu}). In this context, those sources that have jets detected toward them 
are very young (well before the MS turn-on), while those without detectable jets near the protostar have ultracompact HII (UCHII) 
regions. If the disappearance of a collimated jet in early B protostars is due to the presence of enhanced ionising radiation from an 
accreting early main sequence star, then all early B stars may be formed via accretion. In this sense, they are scaled-up 
versions of low-mass protostars at early phases (as for IRAS\,20126+4104, IRAS\,18151-1208, IRAS\,11101-5829, etc...).
As YSOs evolve, developing UCHII regions that destroy the disc, the jets finally disappear, and the outflows
look more like poorly collimated wind-blow bubbles (see also Arce et al~\cite{arce}).

In conclusion, the three high-mass protostellar jets spectroscopically studied appear as scaled-up versions of the
low-mass ones. Furthermore, a morphological analysis of the few intermediate- high-mass H$_2$ jets known up to now
partially supports an evolutionary jet/outflow scenario. 
A larger sample of intermediate-/high-mass jets is, however, needed, and it would be premature jumping to the conclusion 
that the disc accretion-ejection paradigm can be extended to the intermediate- and high-mass protostars.

\section{Conclusions}
\label{conclusion:sec}
The IRAS\,20126+4104 H$_2$ jet has been extensively investigated through near-IR H$_2$ and [\ion{Fe}{ii}] narrow-band imaging,  
H$_2$ high resolution spectroscopy, along with low resolution spectroscopy (0.9-200\,$\mu$m) throughout the infrared wavelength
range. The kinematical, dynamical, and physical conditions of the H$_2$ gas along the flow have been probed. 
The main results of this work are the following:
\begin{enumerate}

\item[-] A high angular resolution H$_2$ continuum subtracted image from Subaru reveals a small-scale precession of the jet close to the source, with an angle of ($\sim$7\fdg6) and a period of $\sim$1100\,yr. This is about a factor 50 shorter than the precession period deduced from large-scale H$_2$ images.

\item[-] H$_2$ and [\ion{Fe}{ii}] narrow-band images show the appearance of a new knot, labelled X, 
roughly 1$\arcsec$ from the source position. No further ionic emission is detected along the flow in the narrow-band imaging, 
indicating that the jet is mainly molecular. 

\item[-] Low resolution spectra are rich in H$_2$ emission, and no ionic emission is detected along the flow, with the exception 
of faint emission of [\ion{Fe}{ii}] close to the source position (in knots X and B). Faint [\ion{O}{i}] and [\ion{C}{ii}] 
emissions are observed in the ISO-LWS spectrum. They could arise from an embedded PDR region around the source.

\item[-] The peak radial velocities of the knots range from -42 to -14\,km\,s$^{-1}$ in the blue
lobe and from -8 to 47\,km\,s$^{-1}$ in the red lobe. Their line profiles, very broad and often with two or three velocity components,
seem to indicate a bow-shock structure. In both lobes, the absolute peak radial velocities of knots close to the source (A$\cdots$C) are smaller (0--30\,km\,s$^{-1}$) than those located at larger distances (i.\,e. knot D, and knots 1 to 4, 40--50\,km\,s$^{-1}$).
This possibly confirms a change of the flow inclination angle (with respect to the sky) from $\sim$9$\degr$ (close to the source), 
to $\sim$45$\degr$ (in the outer knots). Assuming these inclination values, the spatial velocity of the knots 
is between 50 and 80\,km\,s$^{-1}$.

\item[-] The ro-vibrational diagrams indicate H$_2$ excitation temperatures between 2000 and 2500\,K.
Stratification of temperature is detected only in knot C, which can be modelled combining a warm 
($T_{\rm ex}$=2050\,K) and a hot ($\sim$5200\,K) H$_2$ component. Additionally, ISO-SWS spectrum reveals the presence of a 
cold component (520\,K) with a high column density.

\item[-] Furthermore, our analysis seems to indicate that the H$_2$ is mostly excited in C-type shocks, and no evidence of fluorescent excitation has been observed.

\item[-] The estimated $L_{H_2}$ is 8.1$\pm$0.7\,$L_{\sun}$, where the cold component contributes about 50\% to the whole
radiative cooling. The large H$_2$ luminosity suggests that IRAS\,20126+4104 has a significantly increased accretion rate
compared to the low-mass YSOs. This is also supported by the measured mass flux rates from H$_2$ lines
($\dot{M}_{out}$(H$_2$)$\sim$7.5$\times$10$^{-4}$\,$M_{\sun}$\,yr$^{-1}$) 
well matching the previous CO estimates.
Our analysis also points out that the cold H$_2$ component plays a major role in the kinematics and dynamics of this flow.

\item[-] Comparing the H$_2$ and outflow parameters strongly indicates that the jet is driving, at least partially,
the outflow.

\item[-] By comparing the measured luminosity of the H$_2$ jet with the source bolometric luminosity 
(assumed representative of the accretion luminosity), we show that IRAS\,20126+4104 fits well the correlation between these two quantities already found for low-mass protostellar jets (Caratti o Garatti et al.~\cite{caratti}).

\item[-] Considering our results and literature data of a few intermediate- and high-mass protostellar jets, we conclude that 
these few jets appear to be scaled-up versions of their low-mass protostellar counterparts.
\end{enumerate}

\begin{acknowledgements}

We would like to thank the referee, Debra Shepherd, for her helpful suggestions, which improved the manuscript. 

The present work was supported in part by the European Community's Marie Curie Actions-Human Resource and Mobility within the JETSET (Jet Simulations, Experiments and Theory) network under contract MRTN-CT-2004 005592. Our observations have been funded by the Optical Infrared Coordination network (OPTICON), a major international collaboration supported by the Research Infrastructures Programme of the European Commission's Sixth Framework Programme. 

The version of the ISO data presented in this paper correspond to the Highly Processed Data Product (HPDP) set called 35500738 by W. Frieswijk et al. and HPDP set called 04300333 by Lloyd C., Lerate M. and Grundy T., available for public use in the ISO Data Archive.

The United Kingdom Infrared Telescope is operated by the Joint Astronomy Centre on behalf of the Science and Technology 
Facilities Council of the U.K..

Based in part on observations made with the Italian Telescopio Nazionale Galileo (TNG) operated on the island of La Palma by the Fundaci\'{o}n Galileo Galilei of the INAF (Istituto Nazionale di Astrofisica) at the Spanish Observatorio del Roque de los Muchachos of the Instituto de Astrofisica de Canarias. 

Based in part on data collected at Subaru Telescope and obtained from the SMOKA, which is operated by the Astronomy Data Center, National Astronomical Observatory of Japan.

This research has also made use of NASA's Astrophysics Data System Bibliographic Services and the SIMBAD database, operated
at CDS, Strasbourg, France, and the 2MASS data, obtained as part of the Two Micron All Sky Survey, 
a joint project of the University of Massachusetts and the Infrared Processing and Analysis Center/California Institute of Technology, 
funded by the National Aeronautics and Space Administration and the National Science Foundation.
\end{acknowledgements}

\end{document}